\documentclass{IEEEtran}
\usepackage{tabularx}
\usepackage{subfigure}
\usepackage{stix}
\usepackage{amsmath}
\usepackage{amssymb}
\usepackage{algorithm}
\usepackage{algpseudocode}
\usepackage{graphicx}
\usepackage{color}
\usepackage{latexsym}
\usepackage{epsfig}
\usepackage{url}
\usepackage{hyperref}
\usepackage{soul}
\usepackage{amsthm}
\usepackage[usenames,dvipsnames]{pstricks}
\usepackage{pst-plot}
\usepackage{comment}
\input{Jerry.def}

\newcommand \pd {{\mathsf{PD}}}
\newcommand \pde {{\mathsf{PD}^+}}

\newcommand \WW {{\mathcal{W}}}
\newcommand \WX {{\mathcal{X}}}
\newcommand \WY {{\mathcal{Y}}}
\newtheorem{prop}{Property}
\newcommand \pG {{\tilde{\mathbf{G}}_N}}

\definecolor{darkgreen}{rgb}{0, 0.5, 0}

\begin{document}
\title{Toward Universal Decoding of Binary Linear Block Codes via Enhanced Polar Transformations}

\author{Chien-Ying Lin, Yu-Chih Huang, {\it Senior Member, IEEE,} Shin-Lin Shieh, {\it Senior Member, IEEE,}\\ and Po-Ning Chen {\it Senior Member, IEEE}


\thanks{C.-Y. Lin, Y.-C. Huang, and P.-N. Chen are with the Institute of Communications Engineering, National Yang Ming Chiao Tung University, Hsinchu 300093, Taiwan (email: thisistt.c, jerryhuang, poningchen@nycu.edu.tw).}

\thanks{S.-L. Shieh is with the Department of Electrical Engineering and the Institute of Communications Engineering, National Tsing Hua University, Hsinchu 300044, Taiwan (e-mail: slshieh@ee.nthu.edu.tw).}
}
\maketitle

\begin{abstract}
Binary linear block codes (BLBCs) are essential to modern communication, {\black but their diverse structures often require tailor-made decoders}, increasing complexity. This work introduces enhanced polar decoding ($\pde$), a universal soft decoding algorithm that transforms any BLBC into a polar-like code compatible with efficient polar code decoders such as successive cancellation list (SCL) decoding. Key innovations in $\pde$ include pruning polar kernels, shortening codes, and leveraging a simulated annealing algorithm to optimize transformations. These enable $\pde$ to achieve competitive or superior performance to state-of-the-art algorithms like OSD and GRAND across various codes, including extended BCH, extended Golay, and binary quadratic residue codes, with significantly lower complexity. Moreover, $\pde$ is designed to be forward-compatible with advancements in polar code decoding techniques and AI-driven search methods, making it a robust and versatile solution for universal BLBC decoding in both present and future systems.
\end{abstract}

\section{Introduction}
Binary linear block codes (BLBC), a significant subclass of error correction codes, have been widely used in modern communication systems \cite{shulin09}. Since Hamming's pioneering work \cite{Hamming1950}, numerous types of codes have been developed, each with unique strengths. For instance, BCH codes \cite{shulin09} are renowned for their large minimum Hamming distance, low-density parity-check (LDPC) codes \cite{Gallager62} enable low-complexity soft decoding and provide excellent finite-length performance, and polar codes \cite{arikan09} are celebrated for being asymptotically capacity-achieving and free from error floors \cite{Mondelli16polar_floor}. The diverse strengths of these codes have led to the implementation of multiple types of codes within a single system. A recent example is the 5G cellular system \cite{5G_book18, Jinhong_6Gcodes}, where polar codes and LDPC codes are employed as the coding techniques for the control and data channels, respectively. {\black To decode those codes, the current approach is to equip each device with multiple decoders, one for each type of code, which necessarily increases the decoding complexity and decoder size. This motivates the pursuit of universal decoding, namely, a low-complexity decoding algorithm capable of decoding {\it all} types of codes. In particular, we consider a {\it parametrized} universal decoder, which enables multiple codes to share the same decoding circuit; only a small set of parameters needs to be stored and applied, greatly reducing hardware overhead.}


Another compelling motivation for pursuing universal decoding is the challenge of decoding powerful algebraic codes. Algebraic codes form a substantial subset of BLBC, utilizing algebraic structures to achieve excellent Hamming distances and robust error-correction capabilities \cite{blahut2003algebraic}. For instance, BCH codes are widely employed in disk storage, CDs, and satellite communications due to their strong multi-error correction capability. Similarly, Golay codes, small perfect error-correcting codes, are used in applications like deep space and radio communications. However, despite their many advantages, most algebraic codes face a limitation: they are inherently difficult to soft-decode \cite{shulin09}.\footnote{While soft-decoding methods for algebraic codes do exist, see for example \cite{Krishna08, Li_Chen22}, they remain complex and are not easily generalizable to other codes.} As a result, efficient and low-complexity soft-decoding algorithms for algebraic codes are still in demand. Developing a low-complexity universal soft-decoding approach could effectively address this challenge.

While universal decoding techniques do exist, each comes with its own limitations. Maximum likelihood decoding (MLD) is perhaps the earliest universal decoding method, offering optimal performance for decoding any BLBC. However, MLD requires examining every possible codeword to make a final decision, causing its complexity to grow exponentially with the code dimension. Ordered statistics decoding (OSD) \cite{OSD} is a code-agnostic, soft-decision decoding algorithm that approximates optimal decoding by reordering the reliability of the received bits. Since its invention, OSD has been widely adopted for decoding various codes, demonstrating near-optimal performance. However, its computational complexity increases significantly with the code length and the decoding order. More recently, a universal decoding algorithm called guessing random additive noise decoding (GRAND) \cite{GRAND,ORBGRAND} was introduced. Unlike traditional methods, GRAND recovers the original codeword by guessing the random noise added during transmission rather than the transmitted codeword itself. GRAND exhibits excellent performance for various high-rate codes; however, to the best of our knowledge, its effectiveness for low-rate codes remains largely unexplored.


In \cite{PD}, we proposed a universal decoding algorithm named polar decoding ($\pd$). The core idea of $\pd$ is to transform a BLBC into a polar code with dynamic frozen bits \cite{Trifonov13} and decode it using a polar code decoding algorithm. This is of enormous practical importance, as the implementation of decoding algorithms for polar codes has become highly mature \cite{Fast_SCL, Fast_flex_SCL, Memory_eff_SC, Rowshan22rewind}, thanks to their adoption in 5G. The transformation is achieved through a permutation operation, where different permutations result in polar codes with different sets of dynamic frozen bits. A brute-force search was then used to identify a permutation that yields good decoding performance. In \cite{PD}, we demonstrated that $\pd$ achieves near-MLD performance with lower decoding complexity than OSD for various extended BCH (eBCH) codes and an extended Golay (eGolay) code \cite{Golay2412}. However, a challenging case was also identified in \cite{PD}, pointing out the limitations of $\pd$. Specifically, the following generator matrix was considered:
\begin{equation}\label{eqn:challenging_case}
   \mathbf{G}= \begin{bmatrix}
  1 & 1 & 1 & 0 & 0 & 0 & 0 & 0 \\
  0 & 0 & 0 & 1 & 1 & 1 & 0 & 0 \\
  0 & 0 & 0 & 0 & 0 & 0 & 1 & 1 \\
\end{bmatrix}.
\end{equation}
For this code, it was shown that $\pd$ performs poorly for all $8! = 40320$ permutation matrices. 

In this work, motivated by the insufficiency of $\pd$ highlighted by the above challenging case, we continue our pursuit of low-complexity universal soft decoding. We propose a novel universal soft decoding algorithm called enhanced polar decoding, $\pde$. Unlike $\pd$, which transforms a BLBC into a polar code with dynamic frozen bits, the key idea behind $\pde$ is to transform a BLBC into another type of code. While this transformed code is not strictly a polar code, it remains compatible with polar code decoding algorithms, such as successive cancellation list (SCL) decoding \cite{Tal15}. The key ingredients of the proposed $\pde$ include:
\begin{itemize}
    \item In $\pd$, the set of kernels (multi-kernel polar codes \cite{MultiKernel} are allowed) for the polar codes is first determined, and then a suitable permutation matrix is searched for to transform the target BLBC into a polar code with that set of kernels. This restricts the transformed code to the strict family of polar codes whose generator matrices admit a Kronecker product structure. Moreover, it is challenging to decide which set of kernels to start with and in what order to arrange them. For example, for a code of length 24 that can be transformed into polar codes with kernels of sizes $3\times 2\times 2\times 2$, it is difficult to determine where to place the kernel of size 3 and which kernel of size 3 to adopt. In $\pde$, we propose a novel transformation that converts a BLBC into a polar code with a {\it pruned kernel}. In this approach, we always begin with a polar code using {\black Ar\i kan's} kernel and remove some edges to form a polar code with pruned kernels. The question of which edges to prune is then formulated as a local search problem.

    \item Originally, in $\pd$, the multi-kernel technique is adopted to control the code length. However, for $\pde$, as mentioned above, we adopt the pruning technique to modify {\black Ar\i kan's} kernel. Consequently, another technique is required to control the code length. The second ingredient in the proposed $\pde$ is to further shorten the transformed polar code, tailoring it to meet the desired code length. The problem of determining which bits to shorten is again formulated as a local search problem.

    \item We now face a large local search problem aimed at finding: 1) the permutation matrix, 2) the pruning pattern, and 3) the shortening pattern. The third ingredient of the proposed approach is to solve this local search problem with an AI-inspired search algorithm. Specifically, we develop a simulated annealing algorithm \cite{simu_anneal} to solve this extensive local search problem.
\end{itemize}

In summary, the main contribution of this work is the proposal of a novel universal decoding algorithm, $\pde$, which transforms a BLBC into a polar-like code. After transformation, this code becomes a shortened version of a polar code with pruned {\black Ar\i kan's} kernels and dynamic frozen bits. {\black As a result, it can still be decoded by a decoding algorithm for polar codes.} The algorithm is universal in the sense that it can decode \emph{any} BLBC, while its performance depends on the quality of the search results. Extensive simulations presented in this paper demonstrate that, for many codes—including eBCH codes, eGolay codes, and binary quadratic residue codes—$\pde$ together with SCL decoding achieves performance that is better than or comparable to OSD, GRAND, and $\pd$, while offering significantly lower decoding complexity.

Another critical advantage of $\pde$ that deserves emphasis is its {\it forward compatibility}, which manifests in the following two aspects: 1) The proposed $\pde$ piggybacks on polar codes, enabling it to benefit from any existing or future advancements in decoding algorithms for polar codes. This makes $\pde$ highly adaptable and forward-compatible with ongoing progress in polar code research. In this work, we adopt SCL decoding for polar codes; however, any other decoding algorithm, such as SCL flip decoding \cite{viterbo_SCLflip}, BP decoding \cite{BP_polar}, BP list decoding \cite{BP_list}, sequential decoding \cite{Seq_dec_polar}, automorphism ensemble decoding \cite{Automorphism_Ensemble_Dec}, SC ordered search decoding \cite{PeihongYuan24}, etc, can also be utilized. This forward compatibility is immensely valuable, as polar codes have been central to coding research for many years. Numerous efficient and low-complexity decoding algorithms have been developed—and will continue to be developed—for polar codes. By leveraging polar codes, $\pde$ can reap substantial benefits from these advancements. 2) The proposed $\pde$ heavily relies on solving a large-scale search problem, which is currently tackled using simulated annealing. This, too, is forward-compatible with any future developments in local search methods or AI algorithms that could provide improved solutions.

\subsection{Organization}
The paper is organized as follows. In Section~\ref{sec:background}, we introduce background knowledge and state the problem. In Section~\ref{sec:proposed}, we discuss the proposed $\pde$ decoding for BLBCs. In Section~\ref{sec:search}, we describe the proposed AI-inspired search algorithm for finding good instances of the proposed $\pde$. Simulation results are provided in Section~\ref{sec:simulation} to validate the proposed decoding algorithm. Some concluding remarks are given in Section~\ref{sec:conclude}.

\section{Background}\label{sec:background}

In this section, we first formulate the problem of decoding BLBCs in Section~\ref{subsec:problem}. Next, we briefly review the conventional polar codes, multi-kernel polar codes, and polar codes with dynamic frozen bits in Section~\ref{subsec:polar}. $\pd$ of BLBC is then reviewed in Section~\ref{subsec:PD_review}.

\subsection{Decoding of BLBC}\label{subsec:problem}
Let $\mathcal{C}$ be an {\black $(n,k)$-BLBC} that encodes a $k$-bit message $\mathbf{m}\in\mathbb{F}_2^k$ into a $n$-bit codeword $\mathbf{c}\in\mathbb{F}_2^n$ whose relationship is given by
\begin{equation}
    \mathbf{c}=\mathbf{m}\mathbf{G},
\end{equation}
where $\mathbf{G}\in\mathbb{F}_2^{k\times n}$ is a generator matrix. A modulated codeword $\mathbf{x}$ is sent to the channel $\mathcal{W}:\WX\rightarrow\WY$ and a noisy output $\mathbf{y}$ arrives at the receiver. A decoding function $\mathsf{dec}$ is applied on $\mathbf{r}$ a post-processed version of $\mathbf{y}$ to an estimate $\hat{\mathbf{m}}=\mathsf{dec}(\mathbf{r})$.
The goal of this paper is to design a $\mathsf{dec}$ that can provide low $p_e$ for a large class of codes $\mathcal{C}$.

\subsection{Conventional Polar Codes}\label{subsec:polar}
Polar codes, introduced by Erdal {\black Ar\i kan} in 2009 \cite{arikan09}, represent the first family of capacity-achieving codes for binary memoryless channels, featuring explicit construction and low encoding/decoding complexity. What follows is a brief review of polar codes.


To facilitate the discussion, we use the binary memoryless symmetric (BMS) channel as the channel model. Let $\WW: \WX \rightarrow \WY$ represent a generic BMS channel. Here, $\WX=\{0,1\}$ is the input alphabet, and $\WY$ is the output alphabet of the channel. The probability of observing $y \in \WY$ given that $x \in \WX$ was transmitted is denoted by $\WW(y|x)$. For using channel $n$ times, due to the memoryless property, we have $\WW(\mathbf{y}|\mathbf{x})=\prod_{i=1}^n \WW(y_i|x_i)$. To profile the channel performance, the Bhattacharyya parameter\footnote{Another popular way of measuring the quality of the channel in the polar coding literature is the mutual information. Here, since our focus is on reducing $p_e$, rather than achieving higher rates, we consider solely the Bhattacharyya parameter.} is usually considered as a measure of the reliability, which is given by
\begin{align}
Z(\WW) \triangleq \sum_{y \in \WY} \sqrt{\WW(y|0)\WW(y|1)}.
\end{align}

\begin{figure}[tbh]
\centering
\includegraphics[width=0.25\textwidth]{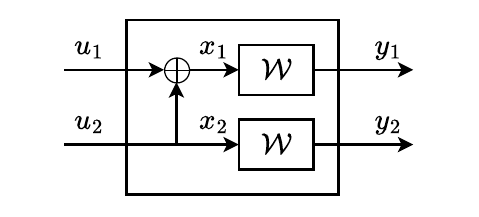}
\caption{{\black Ar\i kan's} kernel.}
\label{fig:arikan_kernel}
\end{figure}

We start by explaining the channel polarization for $n=2$. Consider the basic polarization kernel (or {\black Ar\i kan's} kernel) as shown in Fig.~\ref{fig:arikan_kernel}, which transforms $(u_1,u_2)$ into $(x_1,x_2)$ by $x_1 = u_1 \oplus u_2$ and $x_2 = u_2$. i.e., $\mathbf{x}=\mathbf{u}\mathbf{G}_2$, where
\begin{equation}\label{eqn:Arikan_kernel}
   \mathbf{G}_2=\mathbf{F}_2= \begin{bmatrix}
  1 & 0 \\
  1 & 1 \\
\end{bmatrix}.
\end{equation}



At the receiver, two new channels are formed. For the first one, we decode $u_1$ by $(y_1,y_2)$, giving rise to $\WW^-: \WX \rightarrow \WY^2$, which we abbreviated by the operation {\black $\boxast$. i.e., $\WW^- = \WW \boxast \WW$.} For the second one, we decode $u_2$ by $(y_1,y_2,u_1)$, giving rise to $\WW^+: \WX \rightarrow \WY^2 \times \WX$, represented by the operation {\black $\circledast$. Thus, $\WW^+ = \WW \circledast \WW$.} Their transition probabilities are given as follows:
\begin{equation}
(\WW \boxast \WW)(y_1, y_2 | u_1) =
\frac{1}{2} \sum_{u_2 \in \WX} \WW(y_1 | u_1 \oplus u_2) \WW(y_2 | u_2),
\end{equation}
and
\begin{equation}
(\WW \circledast \WW)(y_1, y_2, u_1 | u_2) = \frac{1}{2}\WW(y_1 | u_1 \oplus u_2) \WW(y_2 | u_2),
\end{equation}
respectively.


It was shown in \cite{arikan09} that the channel is polarized in the sense that 
\begin{align}
Z(\WW^-) &\leq 2Z(\WW)- Z(\WW)^2, \\
Z(\WW^+) &= Z(\WW)^2,
\end{align}
and
\begin{align}
Z(\WW^-) \leq Z(\WW) \leq Z(\WW^+).
\end{align}
To polarize a channel with $n=2^m$ channel uses for a positive integer $m$, the above process can be employed $m$ times recursively. For example, for the $\WW^{2N}$ channel with $N\in\{1,2,\ldots, 2^m\}$, we have
\begin{align}
&\WW^{2N}_{2i-1} (y_1^{2N}, u_1^{2i-2} | u_{2i-1})=\WW^N \boxast \WW^N \nonumber \\
&\hspace{10pt}=\sum_{u_{2i}} \frac{1}{2} \sum_{u_{2i+1,e}^{2N}} \frac{1}{2^{N-1}} \WW^{N} (y_{N+1}^{2N} | u_{1,e}^{2N}) \nonumber \\
&\hspace{10pt}\times \sum_{u_{2i+1,o}^{2N}} \frac{1}{2^{N-1}} \WW^{N} (y_1^N | u_{1,o}^{2N} \oplus u_{1,e}^{2N} ),
\end{align}
and 
\begin{align}
&\WW^{2N}_{2i} (y_1^{2N}, u_1^{2i-1} | u_{2i}) = \WW^N \circledast \WW^N
\nonumber \\
&\hspace{10pt}=\frac{1}{2} \sum_{u_{2i+1,e}^{2N}} \frac{1}{2^{N-1}} \WW^{N} (y_{N+1}^{2N} | u_{1,e}^{2N}) \nonumber \\
&\hspace{10pt}\times \sum_{u_{2i+1,o}^{2N}} \frac{1}{2^{N-1}} \WW^{N} (y_1^N | u_{1,o}^{2N} \oplus u_{1,e}^{2N} ).
\end{align}

\begin{figure}[tbh]
\centering
\includegraphics[width=0.50\textwidth]{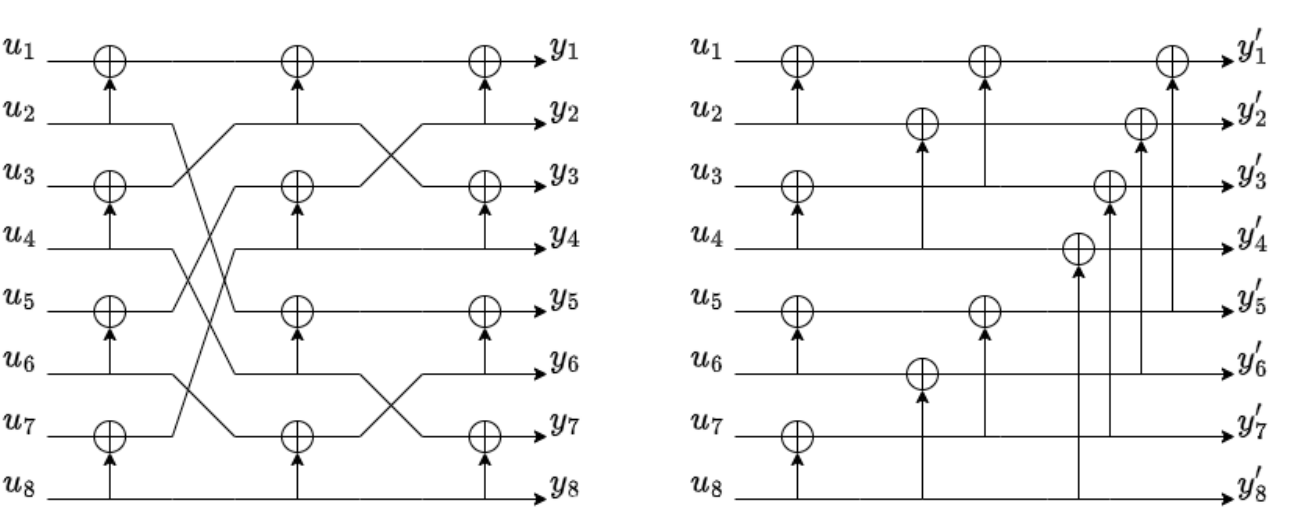}
\caption{LHS: Recursive structure of {\black Ar\i kan's} polarization with $n=8$. RHS: The bit-reversed version of the same structure. }
\label{fig:polar_example_len8}
\end{figure}

The generator matrix of the conventional polar code can be represented by the $m$th Kronecker product of $\mathbf{F}_2$ as $\mathbf{G}_n=\mathbf{B}_n\mathbf{F}_2^{\otimes m}$, where $\mathbf{B}_n$ is the $n\times n$ bit-reversal permutation matrix. An example of $n=8$ can be found in Fig.~\ref{fig:polar_example_len8}.

To construct an $(n, k)$ polar code for transmitting a $k$-bit message $\mathbf{m}$, the best $k$ channels are identified after channel polarization. The message $\mathbf{m}$ is then encoded as $\mathbf{u} = \pi([\mathbf{m}, \mathbf{0}])$, where $\mathbf{0}$ is an all-zero vector of size $n-k$, and $\pi$ is a permutation operation that maps $\mathbf{m}$ to the best $k$ channels with indices in $\mathcal{I}$. The polar codeword is given by
\begin{equation}\label{eqn:polar_eqn}
    \mathbf{c} = \mathbf{u}\mathbf{G}_n = \pi([\mathbf{m}, \mathbf{0}])\mathbf{G}_n,
\end{equation}
where $\mathbf{G}_n$ is the generator matrix of the polar code. The channels with indices in $\mathcal{I}^c$ are referred to as frozen channels.

In \cite{arikan09}, successive cancellation (SC) decoding is employed to decode polar codes. In SC, 
for decoding each bit $u_i$, the decoder uses previously decoded (and frozen) bits and likelihoods of the received vector $\mathbf{y}$ to estimate the current bit $u_i$. If $i \in \mathcal{I}^c$, $u_i$ is fixed to zero. This process continues until all $n$ bits are decoded, yielding $\hat{\mathbf{c}} = \hat{\mathbf{u}} \mathbf{G}_n$. In \cite{Tal15}, SCL further extends this by maintaining a list of $L$ candidate decoding paths. At each step, the list is expanded by considering both possible values of $u_i$ and retaining the $L$ most likely paths. The final output is the codeword corresponding to the most likely path. 
It was shown in \cite[Proposition 2]{arikan09} that the probability of frame error of a polar code with SC decoding can be upper bounded by the sum of the Bhattacharyya parameters in $\mathcal{I}$ as{\black
\begin{equation}\label{eqn:Bhatt_bound_pe}
    p_e\leq \sum_{i\in\mathcal{I}} Z(W_i^n).
\end{equation}
}
\subsubsection{Other Polarization Kernel and Multi-Kernel Polar Codes}
The asymptotic performance of polar codes has been shown to be tightly connected to the polarization kernel, which is the object undergoing the Kronecker product. For instance, {\black Ar\i kan's} kernel, as shown in \eqref{eqn:Arikan_kernel}, is a size-2 kernel. Kernels with better asymptotic performance than {\black Ar\i kan's} kernel have also been developed; see, for example, \cite{ShuLin_kernel, viterbo20kernel}. Instead of having a length of $2^m$, a polar code with a kernel of size $q$ will have a length of $q^m$.

To construct polar codes of more flexible lengths, multi-kernel polar codes were introduced \cite{MultiKernel}. The main idea is to mix kernels of different sizes while preserving the structure of the Kronecker product and retaining the polarization effect. For example, the generator matrix of a polar code of length 24 can be obtained as $\mathbf{F}_3 \otimes \mathbf{F}_2^{\otimes 3}$, where $\mathbf{F}_3$ is a size-3 kernel.



\subsubsection{Polar codes with Dynamic Frozen Bits}
In conventional polar codes, as described in \eqref{eqn:polar_eqn}, the length-$n$ vector $\mathbf{u}$ is generated by assigning or permuting $\mathbf{m}$ to the good channels via $\pi$, while frozen bits are assigned to the bad channels. It has been observed that codes generated in this manner often have small minimum Hamming distances. However, the frozen bits do not necessarily need to be fixed. In \cite{Trifonov13}, the message $\mathbf{m}$ is first pre-transformed using a $k \times n$ upper-trapezoidal matrix $\mathbf{M}_{\rm DF}$ to form $\mathbf{u}$, which is then multiplied by $\mathbf{G}_n$ to produce the codeword. By employing this approach, the frozen bits can be dynamically generated based on the previous message bits at the encoder and the previously decoded bits at the decoder, respectively. Moreover, by judiciously choosing $\mathbf{M}_{\rm DF}$, the overall generator matrix $\mathbf{M}_{\rm DF}\mathbf{G}_n$ may result in a larger minimum Hamming distance than the conventional polar code.

An example is given by
\begin{equation}
\mathbf{M}_{\rm DF}=\left[
\begin{array}{cccccccc}
0 &1 & 0 & 1  & 0 & 0 & 0 & 0 \\
0 & 0 & 1 & 1  & 0 & 0 & 0 & 0 \\
0& 0 & 0 & 0  & 1 & 0 & 0 & 0
\end{array}
\right],
\end{equation}
where the all-zero columns in columns 1, 6, 7, and 8 correspond to the traditional fixed frozen bits, the standard unit vectors in columns 2, 3, and 5 correspond to the information bits, and the 4-th column corresponds to the dynamic frozen bit that can be dynamically generated by $u_2\oplus u_3$.

\subsection{Review of Polar Decoding}\label{subsec:PD_review}

In \cite{PD}, a universal decoding algorithm, namely $\pd$, was proposed. The main idea of this algorithm is to transform a BLBC into a polar code with dynamic frozen bits and then decode it as such. The transformation is formalized in the following:

\begin{theorem}[Proposition 1 of \cite{PD}]
{\black An $(n,k)$-BLBC} can be transformed into a (possibly multi-kernel) polar code with dynamic frozen bits.
\end{theorem}

An inspection of the proof of the above result reveals how $\pd$ operates as follows:
\begin{align}
\mathcal{C} \triangleq &
\left\{ \mathbf{c} = \mathbf{m} ( \mathbf{E}^{-1} \mathbf{E} ) \mathbf{G} ( \mathbf{P}^{-1} \mathbf{G}_n^{-1} \mathbf{G}_n \mathbf{P} ) \mid     \mathbf{m} \in \mathbb{F}_2^k \right\}, \nonumber \\
= & \left\{ \mathbf{c} = ( \mathbf{m} \mathbf{E}^{-1} ) ( \mathbf{E} \mathbf{G} \mathbf{P}^{-1} \mathbf{G}_n^{-1} ) \mathbf{G}_n \mathbf{P} \mid \mathbf{m} \in \mathbb{F}_2^k \right\} \nonumber \\
= & \left\{ \mathbf{c} = \mathbf{c}_p \mathbf{P} \mid \mathbf{c}_p \in \mathcal{C}_p \right\},
\end{align}
where $\mathbf{G}_n$ is the generator matrix of a polar code of size $n$, $\mathbf{P}$ is an $n \times n$ permutation matrix, {\black $\mathcal{C}_p$ is the polar code corresponding to $\mathbf{G}_n$ and dynmamic fronzen bits induced by $\mathbf{M}_{\rm DF} = \mathbf{E} \mathbf{G} \mathbf{P}^{-1} \mathbf{G}_n^{-1}$,} and $\mathbf{E}$ is the elimination matrix in Gaussian elimination procedure that transforms $\mathbf{G} \mathbf{P}^{-1} \mathbf{G}_n^{-1}$ into upper-trapezoidal form. In $\pd$, by permuting the received signal $\mathbf{y}$ with $\mathbf{P}^{-1}$, the decoding problem\footnote{Since $\mathbf{E}^{-1}$ is always non-singular, decoding $\mathbf{m}$ and $\mathbf{m}\mathbf{E}^{-1}$ are equivalent.} reduces to that of a polar code with dynamic frozen bits induced by $\mathbf{M}_{\rm DF}$.

It is evident that different choices of $\mathbf{P}$ lead to different $\mathbf{M}_{\rm DF} = \mathbf{E} \mathbf{G} \mathbf{P}^{-1} \mathbf{G}_n^{-1}$ and thus different sets of dynamic frozen bits, resulting in variations in decoding performance. A local search problem that finds the best $\mathbf{P}$ minimizing the upper bound on $p_e$ in \eqref{eqn:Bhatt_bound_pe} was formulated and solved by brute-force search in \cite{PD}.



\section{Proposed Enhanced Polar Decoding}\label{sec:proposed}
The proposed $\pde$ builds on the same idea of $\pd$; however, rather than transforming the underlying BLBC to a (possibly multi-kernel) polar code with dynamic frozen bits, we transform it to a shortened version of a polar-like code with dynamic frozen bits that can be decoded by polar code decoding algorithms. In the following subsections, we first discuss the motivation behind the proposed $\pde$ in Section~\ref{subsec:motivation_PDE}. This is followed by a detailed presentation of the main results and the proposed $\pde$ in Section~\ref{subsec:PDE}. The concept of polar-like codes considered in this paper is clarified in Section~\ref{subsec:prune_kernel} through the introduction of pruned kernels. Finally, the shortening operation is detailed in Section~\ref{subsec:shortening}.

\subsection{Motivation}\label{subsec:motivation_PDE}
The proposed $\pde$ is motivated by the drawbacks of $\pd$ listed below:
\subsubsection{Challenging case}
In \cite{PD}, a challenging case was identified, as shown in \eqref{eqn:challenging_case}. It was demonstrated that when $\mathcal{W}$ is an AWGN channel with a raw bit error rate of $0.01$, the sum of the Bhattacharyya parameters in \eqref{eqn:Bhatt_bound_pe} amounts to $0.198997 \times 3 = 0.596991$. However, among the $8! = 40320$ possible choices of permutation matrices, the minimum achievable sum of Bhattacharyya parameters is $0.830539 + 0.149237 + 0.000002 = 0.979778$, which is significantly larger than $0.596991$. This result highlights that $\pd$ is fundamentally inadequate in addressing such cases, and the issue cannot simply be attributed to the large search space.

\subsubsection{Multi-kernel technique to meet the code length}
In $\pd$, the multi-kernel technique was adopted to decode BLBC whose code lengths are not powers of 2. For example, when decoding the $(24,12)$ eGolay code, \cite{PD} considers $\mathbf{G}_{24}=\mathbf{F}_3\otimes\mathbf{F}_2^{\otimes 3}$, where {\black
\begin{equation}
\mathbf{F}_{3}=\left[
\begin{array}{cccccccc}
1 & 1 & 0  \\
1 & 0 & 1  \\
1 & 1 & 1
\end{array}
\right].
\end{equation}
}
However, for a code with an arbitrary block length, it can be challenging to identify suitable kernels and arrange them in an appropriate order to achieve effective polarization.

\subsubsection{Search algorithm}
A brute-force search algorithm is employed in \cite{PD} to find suitable permutation, which is impractical if not impossible for large $n$. 

Motivated by the above drawbacks, we propose $\pde$, which addresses each of these drawbacks.


\subsection{Main Result and Proposed $\pde$}\label{subsec:PDE}
Now, we present the main result: An enhanced version of the $\pd$ decoding algorithm from \cite{PD}. In $\pde$, to decode an $(n,k)$ code with a $k \times n$ generator matrix $\mathbf{G}$, we begin with the $N \times N$ generator matrix $\mathbf{G}_N$ of the traditional polar code of size $N \geq n$, using {\black Ar\i kan's} kernel as defined in \eqref{eqn:Arikan_kernel}. We then prune the connections in $\mathbf{G}_N$ according to the $N \times \log_2(N)$ pruning matrix $\mathbf{R}$, resulting in $\pG = f(\mathbf{R}, \mathbf{G}_N)$ the generator matrix of a pruned polar code $\tilde{\mathcal{C}}_p$. The details of the pruning pattern $\mathbf{R}$ and the pruning function $f$ will be discussed in Section~\ref{subsec:prune_kernel}. We now state the following theorem:

\begin{theorem}
For an {\black $(n, k)$-BLBC} $\mathcal{C}$, an integer $N>n$, an $N\times N$ permutation matrix $\mathbf{P}$, and {\black an $N\times n$} shortening matrix $\mathbf{S}$ containing $n$ {\black columns} of the identity matrix $\mathbf{I}_N$, there exists a polar-like code\footnote{Here, a polar-like code refers to a shortened version of a pruned polar code.} $\tilde{\mathcal{C}}_p$ with dynamic frozen bits, such that the one-to-one correspondence between the codeword $\mathbf{c}$ in $\mathcal{C}$ and the codeword $\mathbf{c}_p$ in $\tilde{\mathcal{C}}_p$ is given by $\mathbf{c} = \mathbf{c}_p \mathbf{P} \mathbf{S}$.
\end{theorem}
\begin{IEEEproof}
\begin{align}
\mathcal{C} \triangleq & \left\{ \mathbf{c} = \mathbf{mG} \mid \mathbf{m} \in \mathbb{F}_2^k \right\} =\left\{ \mathbf{c} = \mathbf{m}\mathbf{I}_k \mathbf{G}\mathbf{I}_N \mid \mathbf{m} \in \mathbb{F}_2^k \right\} \nonumber\\
= & \left\{ \mathbf{c} = \mathbf{m} ( \mathbf{E}^{-1} \mathbf{E} ) \mathbf{G} ( \mathbf{S}^{\dag} \mathbf{P}^{-1} \pG^{-1} \pG \mathbf{P} \mathbf{S}) \mid \mathbf{m} \in \mathbb{F}_2^k \right\} \nonumber\\
= & \left\{ \mathbf{c} = ( \mathbf{m} \mathbf{E}^{-1} ) ( \mathbf{E} \mathbf{G} \mathbf{S}^{\dag} \mathbf{P}^{-1} \pG^{-1} ) \pG \mathbf{P} \mathbf{S} \mid \mathbf{m} \in \mathbb{F}_2^k \right\} \nonumber\\
= & \left\{ \mathbf{c} =  \mathbf{m}_p \mathbf{M}_{\rm DF} \pG  \mathbf{P} \mathbf{S} \mid \mathbf{m}_p \in \mathbb{F}_2^k \right\} \nonumber\\
= & \left\{ \mathbf{c} = \mathbf{c}_p \mathbf{P} \mathbf{S} \mid \mathbf{c}_p \in \tilde{\mathcal{C}}_p \right\},
\label{eqn:proof}
\end{align}
where $\mathbf{E}$ is the elimination matrix that puts $\mathbf{M}_{\rm DF}=\mathbf{E} \mathbf{G} \mathbf{S}^{\dag} \mathbf{P}^{-1} \pG^{-1}$ into upper-trapezoidal form and $\mathbf{c}_p=\mathbf{m}_p \mathbf{M}_{\rm DF} \pG$ is a codeword of a shortened pruned polar code generated by $\pG=f(\mathbf{R},\mathbf{G}_N)$ with dynamic frozen bits determined by $\mathbf{M}_{\rm DF}=\mathbf{E} \mathbf{G} \mathbf{S}^{\dag} \mathbf{P}^{-1} \pG^{-1}$ with $\mathbf{S}^\dag$ being the Moore–Penrose pseudo-inverse. The proof is complete by noting that $\mathbf{P}$ and $\pG$ are permutation and lower-triangular matrices and are thus invertible.
\end{IEEEproof}

\begin{figure*}[tbh]
\centering
\includegraphics[width=0.9\textwidth]{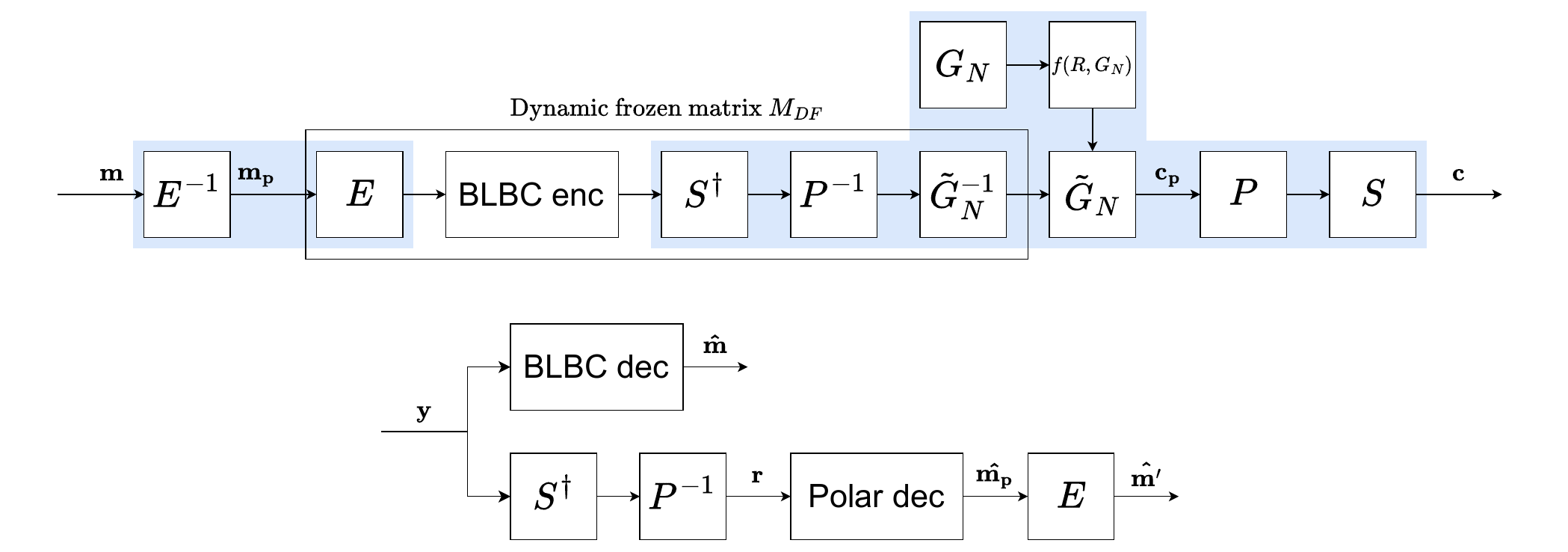}
\caption{A block diagram of the proposed $\pde$ decoder. (Upper) Encoder: All the operations in blue boxes represent virtual components and will not affect the encoding of the original BLBC. The operations inside the black box constitutes $\mathbf{M}_{\rm DF}$. (Lower) Decoder: We can choose either to adopt a decoder for the original BLBC or a decoding algorithm for polar code.}
\label{fig:system}
\end{figure*}

With the above theorem, the proposed $\pde$ at the receiver can treat the code being decoded as $\tilde{\mathcal{C}}_p$. Specifically, we first linearly transform the received signal $\mathbf{y}$ to form
\begin{equation}
\mathbf{r} = \mathbf{y}\mathbf{S}^\dag\mathbf{P}^{-1},
\end{equation}
and interpret it as the received signal corresponding to the output of the channel with input from $\tilde{\mathcal{C}}_p$. Note that although $\tilde{\mathcal{C}}_p$ is pruned, its generator matrix $\pG$ retains the structure of {\black Ar\i kan's} polar code $\mathbf{G}_N$, enabling decoding by polar code algorithms with minor modifications to be discussed in Section~\ref{subsec:prune_kernel}. A block diagram summarizing the decoding process is provided in Fig~\ref{fig:system}.

\begin{remark}\label{rmk:complexity}
    We would like to emphasize that $\mathbf{P}^{-1}$, $\mathbf{S}^\dag$, and $\mathbf{R}$ are all precomputed offline in advance. Consequently, the proposed $\pde$ merely applies these operations. Moreover, $\mathbf{P}^{-1}$ is a permutation matrix, $\mathbf{S}^\dag$ simply adds 0 back,\footnote{\black Since $\mathbf{S}$ is composed of columns of $\mathbf{I}_N$, $\mathbf{S}^\dag$ is simply $\mathbf{S}^T$. In \eqref{eqn:proof}, we write $\mathbf{S}^\dag$ instead of $\mathbf{S}^T$ to emphasize its role in inverting the operation of $\mathbf{S}$.} and $\mathbf{R}$ just disables some edges in $\mathbf{G}_N$, resulting in minimal additional complexity.
    \qed
\end{remark}

\subsection{Pruned Kernel}\label{subsec:prune_kernel}

We now introduce the pruned kernel as a solution to our decoding problem. As mentioned, we always start with a polar code of block length $N > n$ constructed from {\black Ar\i kan's} kernel $\mathbf{F}_2$ as defined in \eqref{eqn:Arikan_kernel}. For {\black Ar\i kan's} kernel $\mathbf{F}_2$, the pruning operation simply removes the edge between $u_1$ and $u_2$ in Fig.~\ref{fig:arikan_kernel}. Consequently, the pruned kernel of size 2 becomes $\tilde{\mathbf{F}}_2 = \mathbf{I}_2$.

Now, since the generator matrix of the polar code of length $N$ is the $m = \log_2(N)$ Kronecker product of {\black Ar\i kan's} kernel, it can be expressed as $\mathbf{G}_N = \mathbf{B}_N \mathbf{F}_2^{\otimes m}$, which can be viewed as a global structure comprising many local {\black Ar\i kan} kernels $\mathbf{F}_2$. In fact, it is obvious that there are $N/2 \cdot \log_2(N)$ such local structures in total. The pruned generator matrix is obtained by pruning some of the local {\black Ar\i kan} kernels to $\tilde{\mathbf{F}}_2 = \mathbf{I}_2$ while leaving others unchanged as $\mathbf{F}_2$.

To represent the pruning pattern, we use an $N/2 \times \log_2(N)$ binary matrix $\mathbf{R}$, {\black where a value of 0 indicates pruning the edge, and 1 indicates no pruning.} Additionally, we denote the pruning operation by $f$ and the generator matrix of the corresponding pruned polar code $\tilde{\mathcal{C}}_p$ by $\pG = f(\mathbf{R}, \mathbf{G}_N)$. An example of the pruned polar code is provided in Example~\ref{ex:prune_8}.

\begin{figure}[tbh]
\centering
\includegraphics[width=0.45\textwidth]{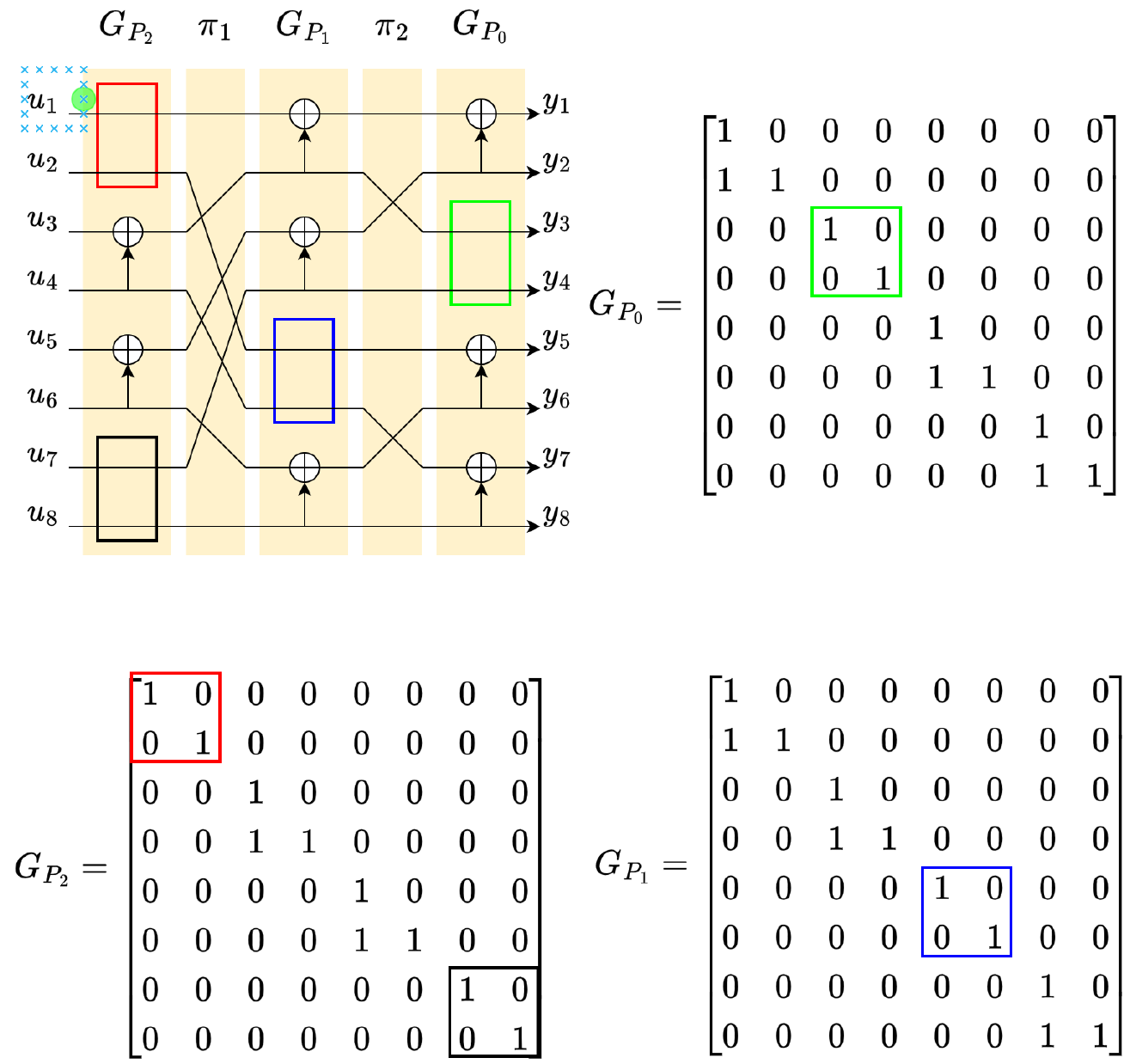}
\caption{An example of a pruned polar code with $N=8$.}
\label{fig:prune_ex}
\end{figure}

\begin{example}\label{ex:prune_8}
    Consider the polar code with $N=8$ constructed with {\black Ar\i kan's} kernel as shown in Fig.~\ref{fig:prune_ex}. In this example, we prune 2 edges in the first stage, 1 edge in the second stage, and 1 edge in the third stage. The corresponding pruning matrix is given by
    \begin{equation}
    \mathbf{R}=\left[ \begin{array}{ccc} \color{red}0 & 1 & 1 \\ [-1.6mm] 1 & 1 & \color{green}0 \\ [-1.6mm]  1 & \color{black}0 & 1 \\ [-1.6mm]  \mathbf{0} & 1 & 1 \end{array}\right],
\end{equation}
where the color codes are used to highlight the corresponding positions in Fig.~\ref{fig:prune_ex}.
\qed
\end{example}

It is worth emphasizing that this pruned kernel retains the butterfly-like structure, ensuring compatibility with both SC and SCL decoders. When a connection is pruned, the data passes through without undergoing polarization. Initially, we have two operations: \( \WW^- = \WW_a \boxast \WW_b \) and \( \WW^+ = \WW_a \circledast \WW_b \). For the pruned kernel, however, we introduce an additional operation, {\black \( \WW_a \mathbin{\square} \WW_b \)}, defined by the following relationships:
\begin{equation}
(\WW_a \mathbin{\square} \WW_b)(y_1, y_2 | u_1) \triangleq  \WW_a(y_1 | u_1),
\end{equation}
and
\begin{equation}
(\WW_a \mathbin{\square} \WW_b)(y_1, y_2, u_1 | u_2) \triangleq  \WW_b(y_2 | u_2).
\end{equation}
These definitions allow SC and SCL decoding algorithms to be modified accordingly to accommodate the pruned kernel.

\begin{remark}\label{rmk:prune}
{\black Prior work has explored polar codes with pruned kernels. The intuition has been that pruning Ar\i kan kernels inherently degrades polar code performance by reducing the mixing crucial for channel polarization. Consequently, pruning has primarily served as a technique to decrease decoding complexity. Take \cite{Wang_prune21} for example, where a kernel is pruned if the two incoming bits are either both linear combinations of message bits or both linear combinations of frozen bits. Consequently, the pruning pattern in the latter stages influences the patterns in earlier stages. Using this idea, the authors of \cite{Wang_prune21} were able to come up with a pruning strategy that constructs polar codes that achieve the capacity for binary erasure channels, while enjoying $\log\log(N)$ per-bit complexity.

In contrast, unlike the construction of standard polar codes, where the selection of the frozen set is a design choice, our problem is dictated by the inherent structure of $(n,k)$-BLBC. Moreover, since $n$ may not always be a power of 2, we are naturally led to consider non-Ar\i kan kernels with sizes that are also not powers of 2. Pruning, in conjunction with shortening, offers a means to generate non-standard kernels with the desired dimensions as illustrated in Figs.~\ref{fig:prune_ex} and \ref{fig:prune_short_43}. Therefore, the intuition that pruning might negatively impact SC decoding performance may not directly apply to our fundamentally different objective of decoding an $(n,k)$-BLBC.
}
\qed
\end{remark}

\subsection{Shortening Operation}\label{subsec:shortening}
We now need to match the blocklength $n$ of the BLBC being decoded. As mentioned, we always start with a polar code whose blocklength $N=2^m$ for some positive integer $m$. In BLBC, there are two simple methods to reduce the codeword length, namely puncturing and shortening. Puncturing involves removing bits from the generated codeword. However, it is difficult to recover punctured bits after transmission through the channel. On the other hand, shortening involves removing bits from the message and generating the same size parity, typically applied in systematic codes where the shortened bits are fixed as zero. Recovering shortened bits is easier because their positions are known, and these shortened bits do not introduce any noise.

Here, we adopt the shortening technique. Specifically, we right-multiply the (permuted) generator matrix $\pG \mathbf{P}$ with the $N \times n$ shortening matrix $\mathbf{S}$, which is composed of selected rows from $\mathbf{I}_N$. It is important to note that this operation constitutes shortening rather than puncturing because $\mathbf{M}_{\rm DF}$ includes the de-shortening matrix $\mathbf{S}^{\dag}$. As a result, the $N-n$ non-selected bit positions are always 0. Hence, when de-shortening is employed at the receiver by adding back 0s to the non-selected bit positions, no information loss is incurred.

\begin{figure*}[tbh]
    \centering
    \includegraphics[width=\textwidth]{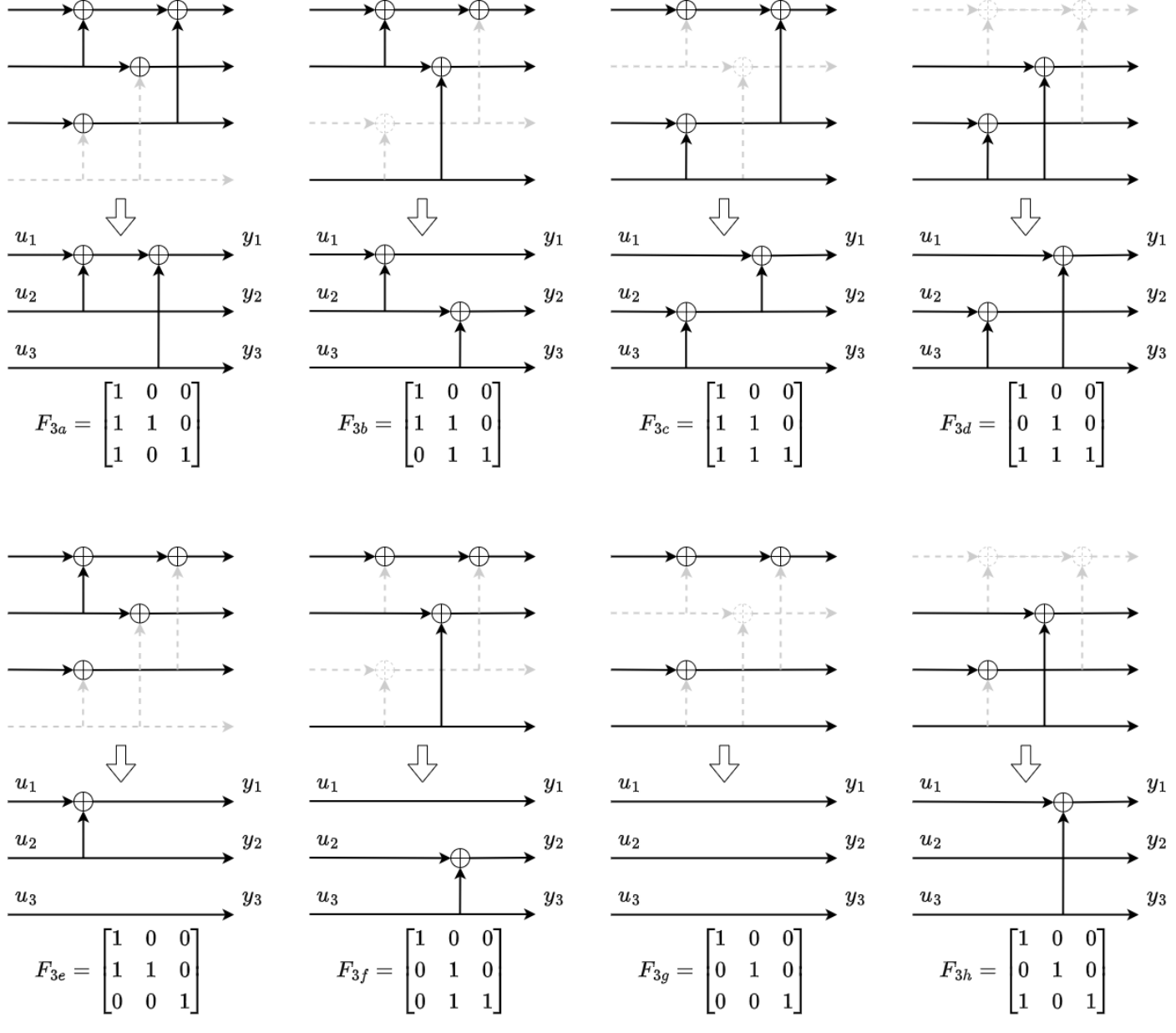}
    \caption{Example of using pruning and shortening to get kernels of size 3 from the {\black Ar\i kan's} kernel.}
    \label{fig:prune_short_43}
\end{figure*}

\begin{remark}
    In this remark, we emphasize that the proposed pruning and shortening operations together enable the generation of a diverse set of kernels. To illustrate this, we consider the example in Fig.~\ref{fig:prune_short_43}. In this example, we begin with a size-4 polar code constructed from \( \mathbf{F}_2 \otimes \mathbf{F}_2 \), demonstrating that all the 8 distinct lower-triangluar kernels of size 3 can be derived through shortening and puncturing. This phenomenon is general, and the diversity of kernels can be significantly expanded by leveraging pruning and shortening, even though we always start with {\black Ar\i kan's} kernels.
    \qed
\end{remark}

\section{AI-Inspired Searching Algorithm}\label{sec:search}
According to the discussion in Section~\ref{sec:proposed}, the proposed method involves three parameters that must be optimized: 1) the permutation operation governed by the permutation matrix $\mathbf{P}$, 2) the pruning operation governed by the pruning function identified as $\mathbf{R}$, and 3) the shortening operation governed by the shortening indices $\mathcal{S}$. {\black The goal is to find $\mathbf{P}$, $\mathbf{R}$, and $\mathcal{S}$ such that the sum of the Bhattacharyya parameters in \eqref{eqn:Bhatt_bound_pe} is minimized. Although this sum serves as an upper bound on SC decoding performance, it is often regarded as a good indicator of the performance of more sophisticated decoders (see, for example, \cite[Conjecture 1]{PeihongYuanConjecture}).}

Before introducing the AI-inspired search algorithm, we first demonstrate that the searches for the permutation and shortening matrices can be combined to reduce complexity.
\begin{lemma}\label{lma:shortening}
For any two shortening patterns $\mathcal{S}_1$ and $\mathcal{S}_2$ with $|\mathcal{S}_1| = |\mathcal{S}_2|$, there exists a permutation matrix $\mathbf{T}$ that transforms one into the other.
\end{lemma}

\begin{IEEEproof}
For {\black $k \in \{1,2\}$,} the shortening operation simply excludes some indices $\mathcal{S}_k \subseteq \{1,\ldots,N\}$ of $\mathbf{c}$. This can be expressed as first forming $\mathbf{c}\mathbf{S}_k$, where $\mathbf{S}_k$ is constructed by setting the columns corresponding to $\mathcal{S}_k$ in the identity matrix $\mathbf{I}_N$ to $\mathbf{0}$, and then transmitting only the indices in $\mathcal{S}_k^c$. Since $\mathcal{S}_k$ is known at both the sender and receiver, the receiver can pad zeros back into the excluded positions and perform decoding as if no shortening had occurred. Given that $|\mathcal{S}_1| = |\mathcal{S}_2|$, it follows that there exists a permutation matrix $\mathbf{T}$ such that $\mathbf{S}_1\mathbf{T} = \mathbf{S}_2$.
\end{IEEEproof}

With Lemma~\ref{lma:shortening}, the search space is reduced to only $\mathbf{P}$ and $\mathbf{R}$, as the shortening pattern can be fixed, for instance, by always shortening the last $N-n$ bits. This effectively incorporates the shortening problem into the search for $\mathbf{P}$ as the received word $\mathbf{y}$ is always multiplied with $\mathbf{P}^{-1}$ in $\pde$. To solve a local search problem, various greedy-like search algorithms exist, such as genetic algorithms, hill climbing algorithms, and simulated annealing. These algorithms differ in their control flow to diverge and converge the candidate pool, yet they share the objective of constructing a candidate-generating function to generate alternative parameters from given ones. In this paper, we consider applying simulated annealing.

In Section~\ref{sec:proposed}, we have identified the pruning pattern as a $\frac{N}{2} \times \log_2{N}$ binary matrix $\mathbf{R}$; each bit acts as a flag to denote a connection status. {\black The permutation matrix can be identified as an one-line notation with length-$N$ integer vector $\mathbf{p}$, where the $i$-th number $p_i$ indicates that we will reorder the $i$-th bit to the $p_i$-th bit.} An example is given in what follows:
\begin{example}
The permutation vector{\black
\begin{equation}
   \mathbf{p}=[1, 5, 3, 7, 2, 6, 4, 8]^T,
\end{equation}}
can be used to equivalently represent the permutatin matrix
\begin{equation}
    \mathbf{P}=\left[
\begin{array}{cccccccc}
1 & 0 & 0 & 0 & 0 & 0 & 0 & 0 \\
0 & 0 & 0 & 0 & 1 & 0 & 0 & 0 \\
0 & 0 & 1 & 0 & 0 & 0 & 0 & 0 \\
0 & 0 & 0 & 0 & 0 & 0 & 1 & 0 \\
0 & 1 & 0 & 0 & 0 & 0 & 0 & 0 \\
0 & 0 & 0 & 0 & 0 & 1 & 0 & 0 \\
0 & 0 & 0 & 1 & 0 & 0 & 0 & 0 \\
0 & 0 & 0 & 0 & 0 & 0 & 0 & 1
\end{array}
\right]
\end{equation}
\qed
\end{example}

The simulated annealing algorithm is a probabilistic optimization technique inspired by the annealing process in metallurgy. It begins with an initial solution \((\mathbf{R}, \mathbf{p})\) and iteratively explores neighboring solutions while progressively narrowing the search space. To generate a feasible neighbor, modifications are made based on the affected component: if changes occur in the pruned kernel part, flipping a single bit produces a feasible neighbor; if changes occur in the permutation matrix part, swapping two elements generates a candidate solution. The candidate-generation process alternates between modifying either the pruned kernel part or the permutation matrix, while keeping the other component fixed.

At each iteration, the simulated annealing algorithm evaluates the fitness of the current solution based on the resulting sum of the Bhattacharyya parameters in \eqref{eqn:Bhatt_bound_pe}, denoted as \(p_{e,\text{current}}\). It then randomly selects a neighboring solution and calculates its fitness, \(p_{e,\text{next}}\). The algorithm accepts the new solution if it has better fitness than the current one. If the new solution is worse, it may still be accepted with a probability given by
\begin{equation}
    \exp\left(-\frac{p_{e,\text{next}} - p_{e,\text{current}}}{T}\right),
\end{equation}
where $T= \gamma^{t-1} T_\text{init}$, for some initial temperature $T_\text{init}$ and $\gamma< 1$,\footnote{In our simulations, we set $T_\text{init}=1$ and $\gamma=0.99999$.} represents the temperature at iteration \(t\). This probabilistic acceptance mechanism allows the algorithm to escape local optima and explore a wider solution space.

A central concept in simulated annealing is the temperature, which controls the likelihood of accepting worse solutions. Initially, the temperature is high, enabling the algorithm to accept worse solutions more frequently to encourage exploration. As the algorithm progresses, the temperature gradually decreases, reducing the probability of accepting worse solutions and making the process increasingly selective. The algorithm continues this iterative process until the maximum number of iterations $t_{\max}$ is met. The final solution is typically either the best solution encountered during the search or the solution found in the last iteration.

\begin{algorithm}
\caption{\textsf{Simulated annealing for searching $(\mathbf{R}^*,\mathbf{p}^*)$}}
\begin{algorithmic}[1]
\State Initialize: $\mathbf{R}_\text{current}$ to be all 1 matrix and {\black $\mathbf{p}_\text{current}=[1,2,\ldots, N]^T$}
\State Initialize: $p_{e,\text{current}}$ using \eqref{eqn:Bhatt_bound_pe} with input $(\mathbf{R}_\text{current},\mathbf{p}_\text{current})$
\State Initialize: $t=1, T_\text{init}=1$
\While{$t\leq t_{\max}$}
    \State $T \leftarrow \gamma^{t-1} T_\text{init}$
        {\black \State Randomly generate $(\mathbf{R}_\text{next},\mathbf{p}_\text{next})$ by either flipping a
        \Statex \hspace{\algorithmicindent}single bit of $\mathbf{R}_\text{current}$ or swapping two elements of
        \Statex \hspace{\algorithmicindent}$\mathbf{p}_\text{current}$}
    \State Calculate $p_{e,\text{next}}$ using \eqref{eqn:Bhatt_bound_pe} with input $(\mathbf{R}_\text{next},\mathbf{p}_\text{next})$
    \If{ $p_{e,\text{next}}<p_{e,\text{current}}$ }
        \State $(\mathbf{R}_\text{current},\mathbf{p}_\text{current})\leftarrow (\mathbf{R}_\text{next},\mathbf{p}_\text{next})$ always
    \Else
        \State {\black $(\mathbf{R}_\text{current},\mathbf{p}_\text{current})\leftarrow (\mathbf{R}_\text{next},\mathbf{p}_\text{next})$ with probability
        \Statex \hspace{\algorithmicindent}\hspace{\algorithmicindent}$\exp(-(p_{e,\text{next}}-p_{e,\text{current}})/T)$}
    \EndIf
\EndWhile
\State \Return $(\mathbf{R}_\text{current},\mathbf{p}_\text{current})$
\end{algorithmic}
\end{algorithm}

\begin{remark}
    Before presenting the simulation results, we would like to reiterate that even with the assistance of AI, we are still tackling a massive local search problem characterized by an astronomically large search space of $2^{\frac{N}{2} \cdot \log_2(N)}N!$. However, as highlighted in Remark~\ref{rmk:complexity}, this search can be performed offline, with the complexity effectively amortized.
    \qed
\end{remark}

\section{Simulation Results}\label{sec:simulation}
In this section, we present simulation results to validate the effectiveness of the proposed $\pde$. For comparison, we also include the performance of several benchmark decoding algorithms: OSD, GRAND,\footnote{For GRAND, we implement the segmented ORB-GRAND in \cite{Rowshan23GRAND} by adapting the computer programs in \url{https://github.com/mohammad-rowshan/Segmented-GRAND}.} $\pd$, and, where feasible, MLD. In the legend, OSD$_\ell$ represents OSD of order $\ell$, while GRAND$_M$ denotes GRAND with $M$ candidate noise sequences. For $\pd$ and $\pde$, the subscripts indicate their respective list sizes in SCL decoding. Beyond performance evaluation, we also compare the computational complexity of the considered schemes. Following the approach in {\black \cite{GRAND, Rowshan23GRAND}}, complexity is measured by the number of candidate sequences the decoder processes.

In Section~\ref{subsec:simu_challenge}, the challenging case shown in \eqref{eqn:challenging_case} is addressed. In Section~\ref{subsec:simu_ebch}, eBCH codes are considered. In Section~\ref{subsec:simu_egolay} is revisited, which is followed by simulations for binary quadratic residue codes in Section~\ref{subsec:simu_bqr}. {\black Last but not least, in Section~\ref{subsec:RLC}, we present simulation results for a randomly generated code to address the potential concern that the proposed $\pde$ is only effective for codes like eBCH, eGolay, and binary QR codes, which possess very strong algebraic structures.}\footnote{All the $(\mathbf{R}, \mathbf{P})$ pairs used in our simulations are available at the following link: \url{https://github.com/thisistt/Enhanced-polar-decoding}.}


\subsection{The Challenging Case}\label{subsec:simu_challenge}
As previously mentioned, in \cite{PD}, we exhausted all $40320$ permutation {\black matrices} for $\pd$ and found that the best sum Bhattacharyya parameter is $0.979778$, which is larger than $0.596991$ obtained by viewing them as independent channels.

\begin{figure}[tbh]
\centering
\includegraphics[width=0.40\textwidth]{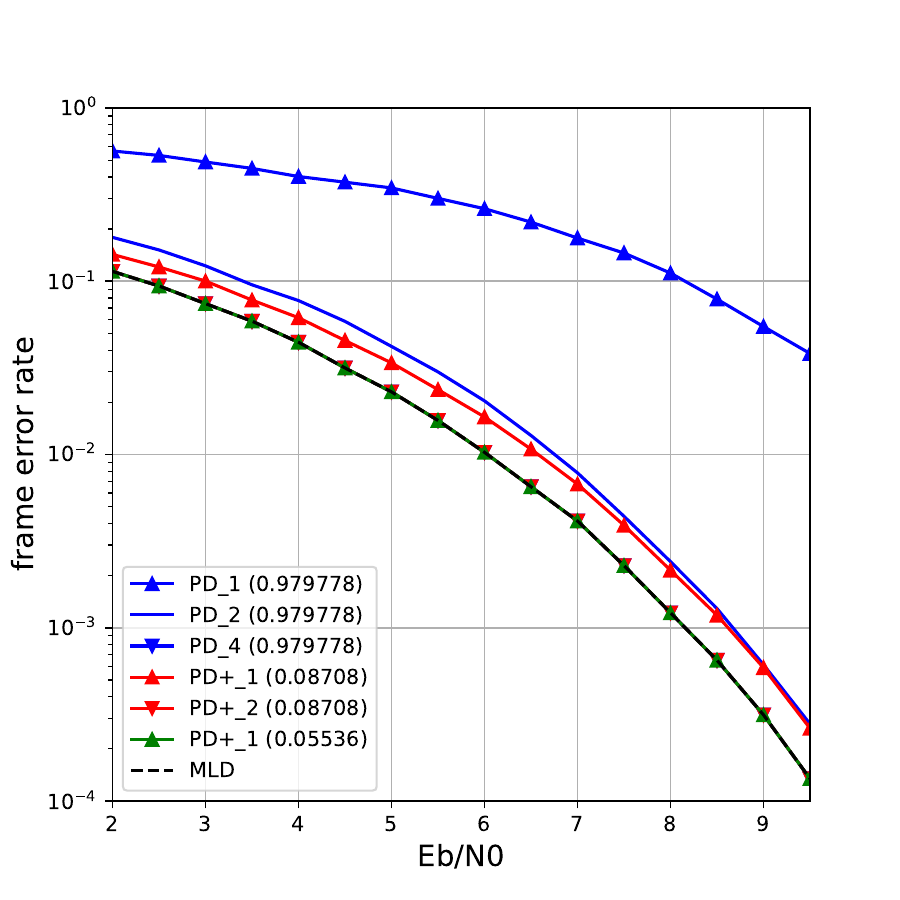}
\caption{Frame error rate vs $E_b/N_0$ for the challenging case. }
\label{fig:Challenging}
\end{figure}

To demonstrate the effectiveness of the proposed $\pde$, we first exhaustively explore all $2^{12}\times 40320\approx 1.6\times 10^8$ candidate transformations and we identify one that achieves a sum Bhattacharyya parameter of $0.05536$, which is only $6\%$ of the value obtained when treating the channels as independent. The simulation results, presented in Fig.~\ref{fig:Challenging}, demonstrate that the newly developed $\pde$ achieves MLD performance with only $L=1$ (i.e., SC decoding), whereas $\pd$ requires $L=4$ to achieve the same performance, which is only marginally less than the effort required to check all 8 codewords, as in MLD.

{\black We also provide the simulation result for the proposed $\pde$ using one of the mediocre transformations, which yields a sum Bhattacharyya parameter of 0.08708. A comparison between the two $\pde$ schemes with $L=1$ supports our choice of using the sum of the Bhattacharyya parameters as the optimization criterion. Moreover, we emphasize that for this challenging case, since all 40320 possible permutations for $\pd$ were exhausted, the results in Fig.~\ref{fig:Challenging} also serve as evidence that incorporating pruning into $\pde$ is crucial for finding improved transformations, as discussed in Remark~\ref{rmk:prune}. }

\begin{figure}[tbh]
\centering
\includegraphics[width=0.40\textwidth]{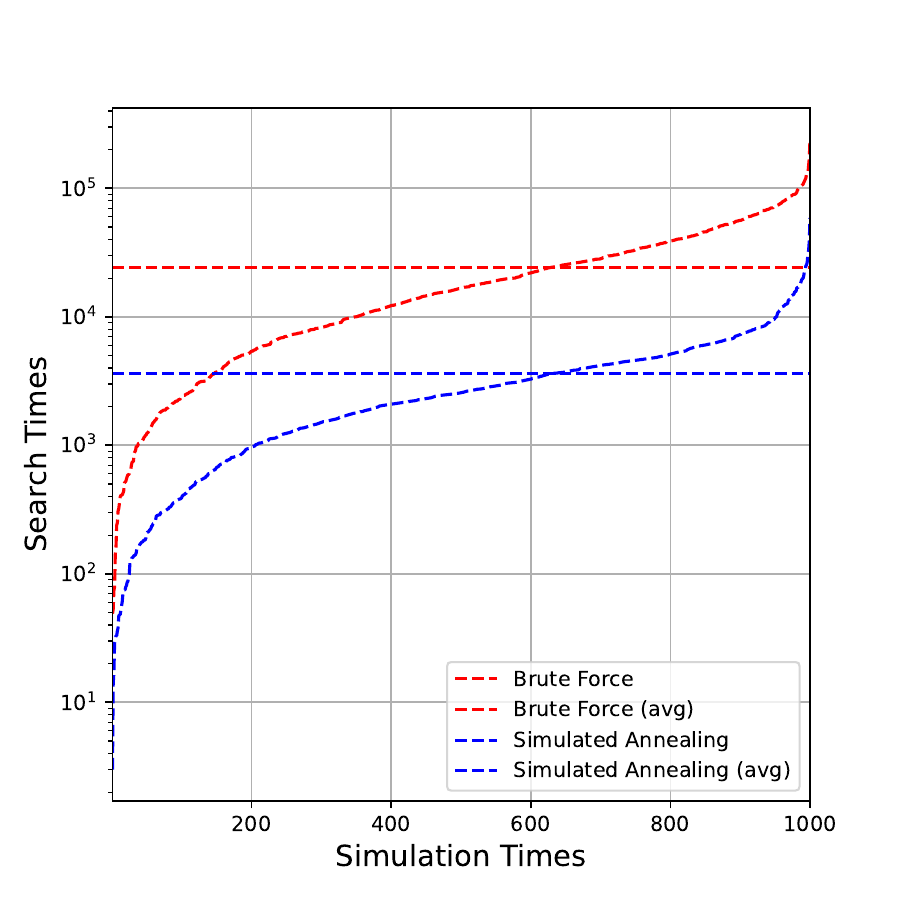}
\caption{Sorted computation time and average computation time.}
\label{fig:compare_search}
\end{figure}

To demonstrate the effectiveness of the proposed simulated annealing search algorithm, we compare its computation time, measured by the number of $(\mathbf{R}, \mathbf{p})$ pairs visited, with that of the exhaustive search for finding the optimal solution. In Fig.~\ref{fig:compare_search}, we present the results of running the search 1,000 times, plotting both the sorted computation times and the average computation time. It can be observed that the proposed simulated annealing algorithm consistently finds the optimal solution while reducing execution time by an order of magnitude. It is important to note that computation time can be significantly reduced by carefully designing the temperature function in simulated annealing or by incorporating advanced techniques from AI/ML. However, these enhancements are beyond the scope of this paper and are not explored further. Additionally, the advantages of an efficient search algorithm, such as the proposed simulated annealing, become even more pronounced for codes of practical lengths, where an exhaustive search is computationally prohibited. In such scenarios, intelligently exploring promising candidates within a limited search time often yields $(\mathbf{R}, \mathbf{p})$ solutions that are substantially better than those obtained through brute-force search.

\subsection{Extended BCH Code}\label{subsec:simu_ebch}
\begin{figure}[tbh]
\centering
\includegraphics[width=0.40\textwidth]{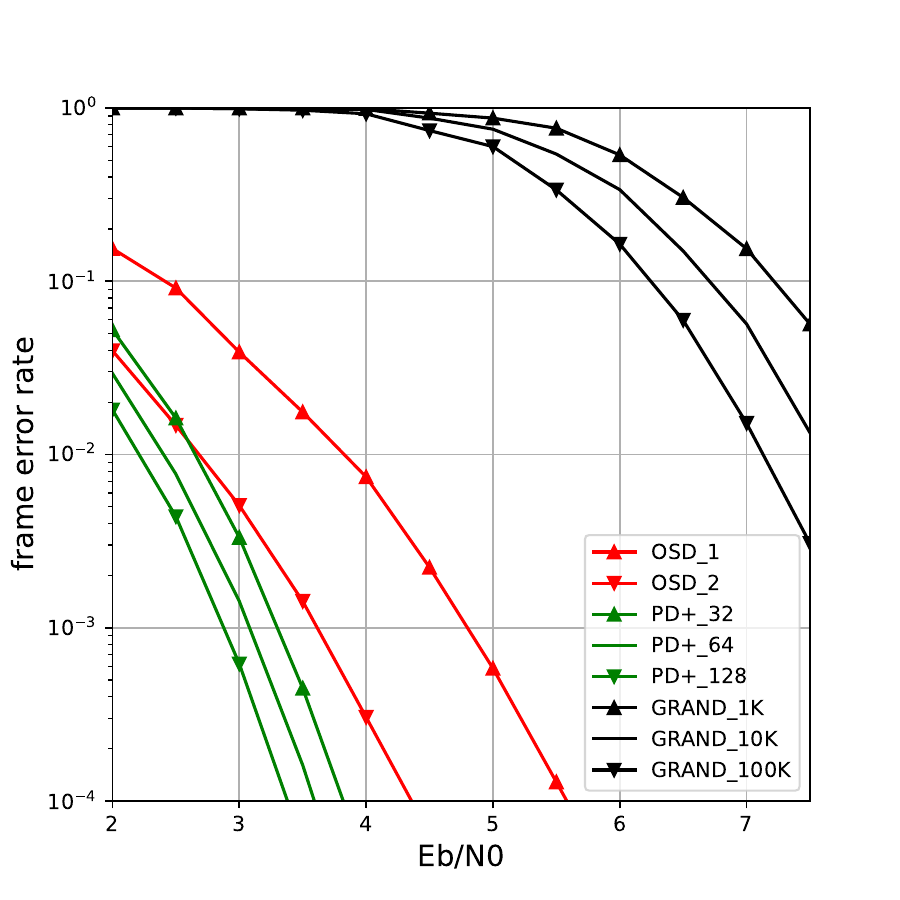}
\caption{Frame error rate vs $E_b/N_0$ for the $(128,57)$ eBCH code.}
\label{fig:eBCH_57}
\end{figure}

We present the results of our proposed $\pde$ for decoding two eBCH codes: the $(128, 54)$ and $(128, 106)$ eBCH codes. Due to the prohibitively high complexity of MLD, its simulation is omitted for these codes. In \cite{PD}, we demonstrated the effectiveness of $\pd$ in decoding eBCH codes with $n=64$. However, for $n=128$, due to its immense computational complexity required by an exhaustive search adopted by $\pd$, we were unable to identify good instances in \cite{PD}. With the introduction of an AI-inspired search algorithm, $\pde$ now enables decoding for codes with $n=128$. The simulation results for the $(128, 57)$ eBCH code are presented in Fig.~\ref{fig:eBCH_57}, with the complexity of each decoding algorithm detailed in Table~\ref{tbl:eBCH_57}. As shown in the figure and table, the proposed $\pde$ with $L=32$ outperforms OSD with order 2 while visiting significantly fewer candidate sequences. In contrast, GRAND struggles to achieve comparable performance even with 100K visited candidates, primarily due to the drastically increased search space associated with $n-k$.

\begin{table}[tbh]
    \centering
    \begin{tabularx}{0.5\textwidth} {
      | >{\centering\arraybackslash}X
      | >{\centering\arraybackslash}X
      | >{\centering\arraybackslash}X
      | >{\centering\arraybackslash}X
      | >{\centering\arraybackslash}X | }
     \hline
    Decoder & OSD$_1$ & OSD$_2$ & $\mathsf{PD}^+_L$ & GRAND$_M$ \\
     \hline
    Candidate & 58 & 1654 & $L$ & $M$ \\
    \hline
    \end{tabularx}
    \vspace{8pt}
    \caption{Complexity comparison for the eBCH $(128, 57)$ case}
    \label{tbl:eBCH_57}
\end{table}

\begin{figure}[tbh]
\centering
\includegraphics[width=0.40\textwidth]{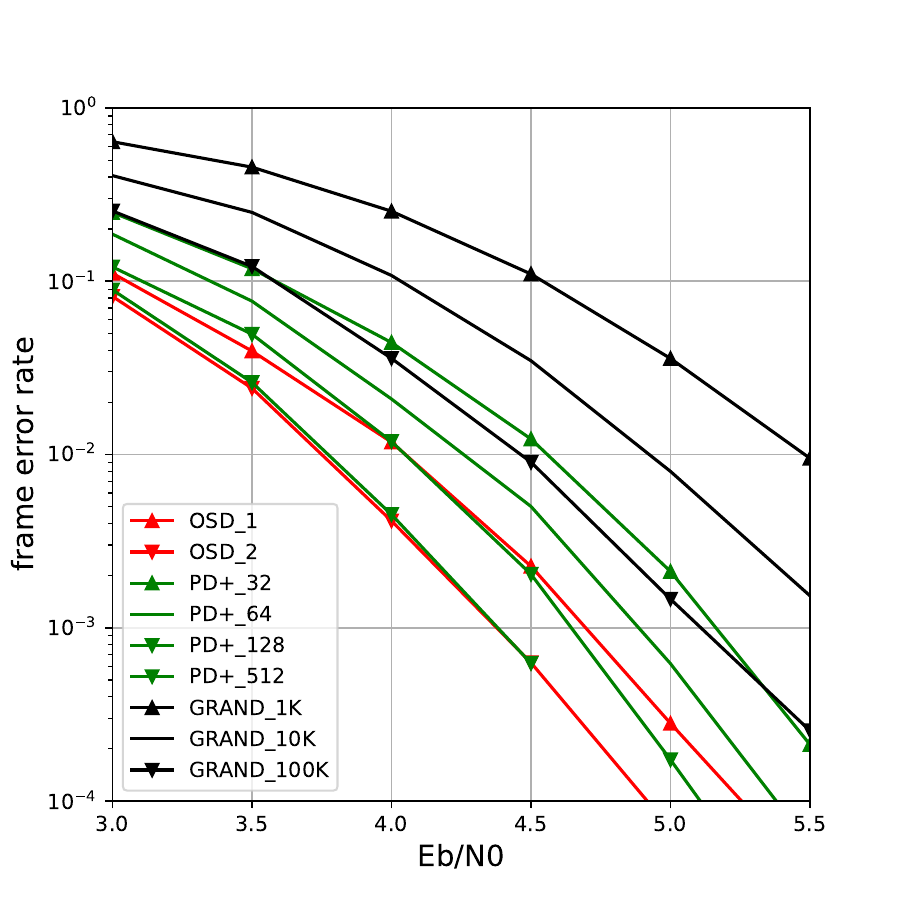}
\caption{Frame error rate vs $E_b/N_0$ for the $(128,106)$ eBCH code.}
\label{fig:eBCH_106}
\end{figure}

In Fig.~\ref{fig:eBCH_106} and Table~\ref{tbl:eBCH_106}, it can be observed that for the $(128, 106)$ eBCH code, the proposed $\pde$ with $L=128$ and that with $L=256$ achieve performance comparable to OSD with order 1 and that with order 2, respectively. Additionally, $\pde$ with $L=32$ demonstrates performance comparable to GRAND with 100K candidate sequences.

It is worth noting that the size of the search space for the proposed $\pde$ is astronomical for the above two codes, at $2^{448} \times 128! \approx 10^{350}$. We believe that a more intelligent search algorithm could be developed to outperform OSD while maintaining significantly lower complexity. We leave this for future work.

\begin{table}[tbh]
    \centering
    \begin{tabularx}{0.5\textwidth} {
      | >{\centering\arraybackslash}X
      | >{\centering\arraybackslash}X
      | >{\centering\arraybackslash}X
      | >{\centering\arraybackslash}X
      | >{\centering\arraybackslash}X | }
     \hline
    Decoder & OSD$_1$ & OSD$_2$ & $\mathsf{PD}^+_L$ & GRAND$_{M}$ \\
     \hline
    Candidate & 107 & 5672 & $L$ & $M$ \\
    \hline
    \end{tabularx}
    \vspace{8pt}
    \caption{Frame error rate vs $E_b/N_0$ for the $(128,106)$ eBCH code}
    \label{tbl:eBCH_106}
\end{table}


\subsection{Extended Golay Code Revisited}\label{subsec:simu_egolay}
\begin{figure}[tbh]
\centering
\includegraphics[width=0.40\textwidth]{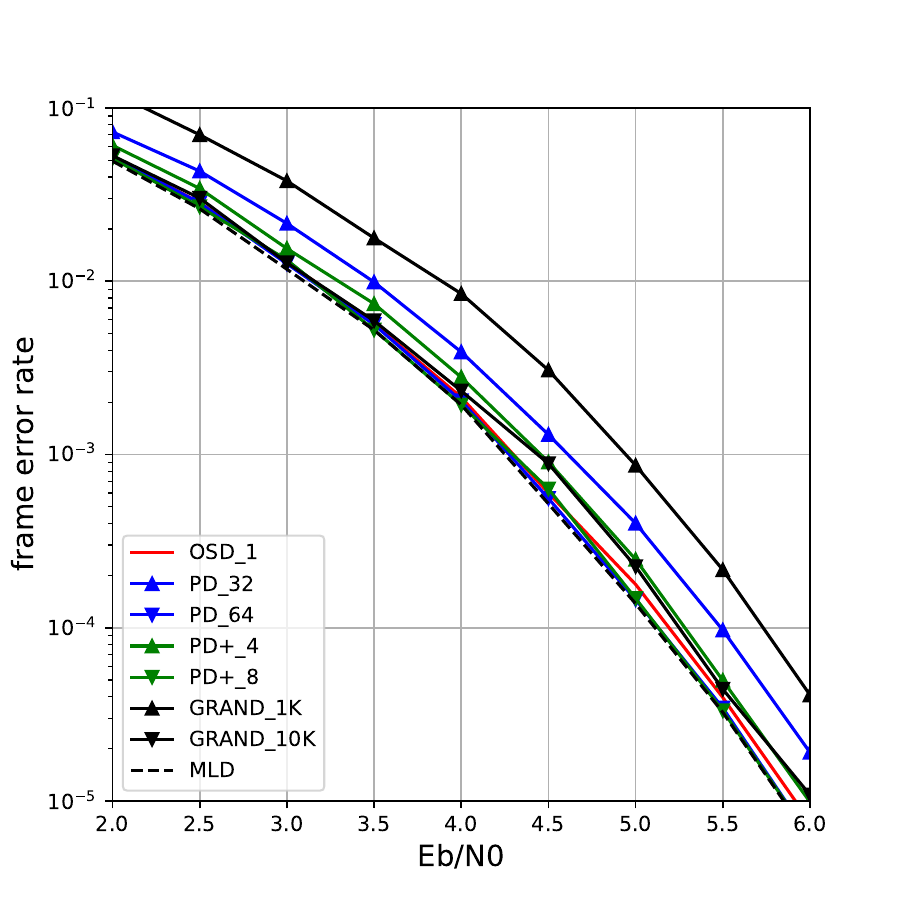}
\caption{Frame error rate vs $E_b/N_0$ for the $(24,12)$ eGolay code.}
\label{fig:eGolay}
\end{figure}

We now revisit the $(24,12)$ eGolay code. In \cite{PD}, we proposed using $\pd$ to transform this code into a multi-kernel polar code with dynamic frozen bits, which is then decoded as a polar code. For the newly proposed $\pde$, we begin with a polar code of size $N=32$ using {\black Ar\i kan's} kernel as defined in \eqref{eqn:Arikan_kernel}. The kernel is subsequently pruned, and the code is shortened to $n=24$. Simulated annealing is employed to identify good pruning and shortening patterns. Our results, presented in Fig.~\ref{fig:eGolay} and Table~\ref{tbl:eGolay}, demonstrate that OSD with order 1, $\pd$ with a list size of 64, GRAND with 10K candidates, and the proposed $\pde$ with a list size of 8 all achieve near-ML performance. Notably, the proposed $\pde$ achieves this performance with the lowest complexity among the compared methods.


\begin{table}[tbh]
    \centering
    \begin{tabularx}{0.5\textwidth} {
      | >{\centering\arraybackslash}X
      | >{\centering\arraybackslash}X
      | >{\centering\arraybackslash}X
      | >{\centering\arraybackslash}X
      | >{\centering\arraybackslash}X | }
     \hline
    Decoder & OSD$_1$ & $\mathsf{PD}_{L'}$ & $\mathsf{PD}^+_L$ & GRAND$_M$ \\
     \hline
    Candidate & 13 & $L'$ & $L$ & $M$ \\
    \hline
    \end{tabularx}
    \vspace{8pt}
    \caption{Complexity in the eGolay $(24, 12)$ Case}
    \label{tbl:eGolay}
\end{table}

\subsection{Binary Quadratic Residue Code}\label{subsec:simu_bqr}
We now focus on the decoding of binary QR codes. Binary QR codes generally have large minimum Hamming distances and are known for achieving the best error performance among binary codes with similar lengths and rates. However, efficiently decoding QR codes is highly challenging, even for hard-decision decoding \cite{QR18}, \cite{QR20}.

\begin{figure}[tbh]
\centering
\includegraphics[width=0.40\textwidth]{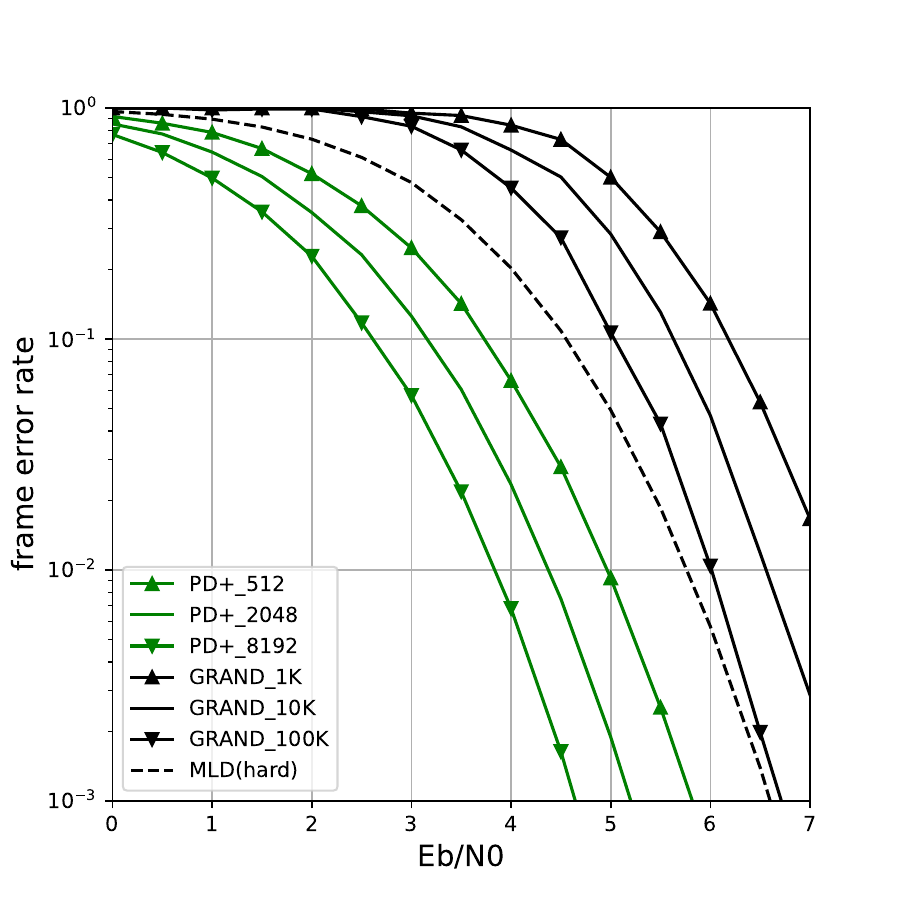}
\caption{Frame error rate vs $E_b/N_0$ for the $(97,49)$ Binary QR code.}
\label{fig:QR_97}
\end{figure}

In Fig.~\ref{fig:QR_97}, we present the simulation results for the $(97,49)$ binary QR code. Notably, 97 is a prime number, which limits $\pd$ in \cite{PD} to starting with a kernel of size 97, which does not polarize. The figure shows that GRAND with 100K candidates achieves performance close to hard-decision ML decoding. For our $\pde$, we start with a polar code of size $N=128$, then prune and shorten it to $n=97$. Simulation results demonstrate that our $\pde$ outperforms hard-decision ML with a list size of $L=128$. Furthermore, the performance of $\pde$ continues to improve as the list size increases. Again, it is worth noting that the search space for this code is astronomical, at $2^{448} \times 97! \approx 10^{286}$. We believe that better instances of $\pde$ could be found, capable of significantly improving performance while requiring a substantially smaller list size.

\begin{table}[tbh]
    \centering
    \begin{tabularx}{0.5\textwidth} {
      | >{\centering\arraybackslash}X
      | >{\centering\arraybackslash}X
      | >{\centering\arraybackslash}X
      | >{\centering\arraybackslash}X | }
     \hline
    Decoder & $\mathsf{PD}_{L'}$ & $\mathsf{PD}^+_L$ & GRAND$_M$ \\
     \hline
    Candidate & $L'$ & $L$ & $M$ \\
    \hline
    \end{tabularx}
    \vspace{8pt}
    \caption{Complexity in the QR $(97, 49)$ Case}
    \label{tbl:QR_97}
\end{table}

{\black
\subsection{A Randomly Generated BLBC}\label{subsec:RLC}
Here, we generate an $8 \times 16$ generator matrix for a systematic encoder, consisting of an $8 \times 8$ identity matrix concatenated with an $8 \times 8$ random binary matrix. The elements of the random matrix are independently drawn from $\mathsf{Bernoulli}(0.5)$. The generator matrix is given by

\begin{equation}
{\scriptsize    \mathbf{G}=\left[
\begin{array}{cccccccccccccccc}
1&0&0&0&0&0&0&0&1&1&0&0&1&0&1&0 \\
0&1&0&0&0&0&0&0&0&0&1&0&1&1&0&0 \\
0&0&1&0&0&0&0&0&1&0&1&0&0&1&0&0 \\
0&0&0&1&0&0&0&0&0&0&1&0&1&0&0&1 \\
0&0&0&0&1&0&0&0&0&1&1&0&1&0&0&0 \\
0&0&0&0&0&1&0&0&0&0&1&0&0&1&0&0 \\
0&0&0&0&0&0&1&0&0&1&0&0&1&0&1&0 \\
0&0&0&0&0&0&0&1&1&0&1&0&0&1&1&1
\end{array}
\right].}
\label{eqn:RLC}
\end{equation}
We input this $\mathbf{G}$ to the proposed $\pde$, and obtain the transformation $(\mathbf{P}, \mathbf{R}, \mathcal{S})$ with $\mathbf{P}$ given in \eqref{eqn:RLC}, $\mathbf{R}$ as illustrated in Fig.~\ref{fig:pruned16}, and $\mathcal{S}=\emptyset$ (i.e., $\mathbf{S}=\mathbf{I}$).

\begin{equation}
    {\scriptsize \mathbf{P}=\left[
\begin{array}{cccccccccccccccc}
0&0&0&0&0&0&0&0&0&0&0&0&0&0&0&1 \\
0&0&0&1&0&0&0&0&0&0&0&0&0&0&0&0 \\
0&0&0&0&0&0&0&0&0&0&0&0&0&1&0&0 \\
0&0&0&0&0&0&0&0&0&1&0&0&0&0&0&0 \\
0&0&0&0&0&0&1&0&0&0&0&0&0&0&0&0 \\
0&0&0&0&0&1&0&0&0&0&0&0&0&0&0&0 \\
0&0&0&0&0&0&0&0&0&0&1&0&0&0&0&0 \\
0&0&0&0&0&0&0&0&0&0&0&0&1&0&0&0 \\
0&0&0&0&0&0&0&1&0&0&0&0&0&0&0&0 \\
0&0&1&0&0&0&0&0&0&0&0&0&0&0&0&0 \\
0&0&0&0&1&0&0&0&0&0&0&0&0&0&0&0 \\
0&1&0&0&0&0&0&0&0&0&0&0&0&0&0&0 \\
0&0&0&0&0&0&0&0&1&0&0&0&0&0&0&0 \\
1&0&0&0&0&0&0&0&0&0&0&0&0&0&0&0 \\
0&0&0&0&0&0&0&0&0&0&0&0&0&0&1&0 \\
0&0&0&0&0&0&0&0&0&0&0&1&0&0&0&0
\end{array}
\right].}
\label{eqn:RLC}
\end{equation}

\begin{figure}[tbh]
\centering
\includegraphics[width=0.5\textwidth]{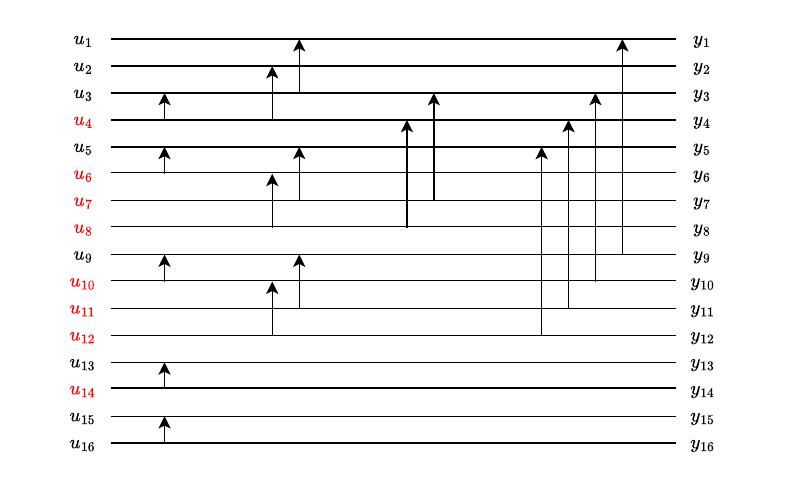}
\caption{The pruned kernel and the polarization result. The variables highlighted in red make up the information set.}
\label{fig:pruned16}
\end{figure}

Simulation results for this randomly generated code, decoded using $\pd$ with $\mathbf{P} = \mathbf{I}$ and the proposed $\pde$, are shown in Fig.~\ref{fig:RLC}. As illustrated in the figure, applying $\pd$ with $\mathbf{P} = \mathbf{I}$ and $L = 1$ results in a performance significantly inferior to that of MLD, indicating that this randomly generated BLBC lacks a polar-like structure. In contrast, the performance greatly improves when the proposed $\pde$ is employed, and they closely approach the MLD performance at $L = 8$. These results further demonstrate the effectiveness of our proposed transformation in converting a BLBC--even one without inherent polar-like structure--into a form amenable to efficient and near-optimal decoding.

\begin{figure}[tbh]
\centering
\includegraphics[width=0.40\textwidth]{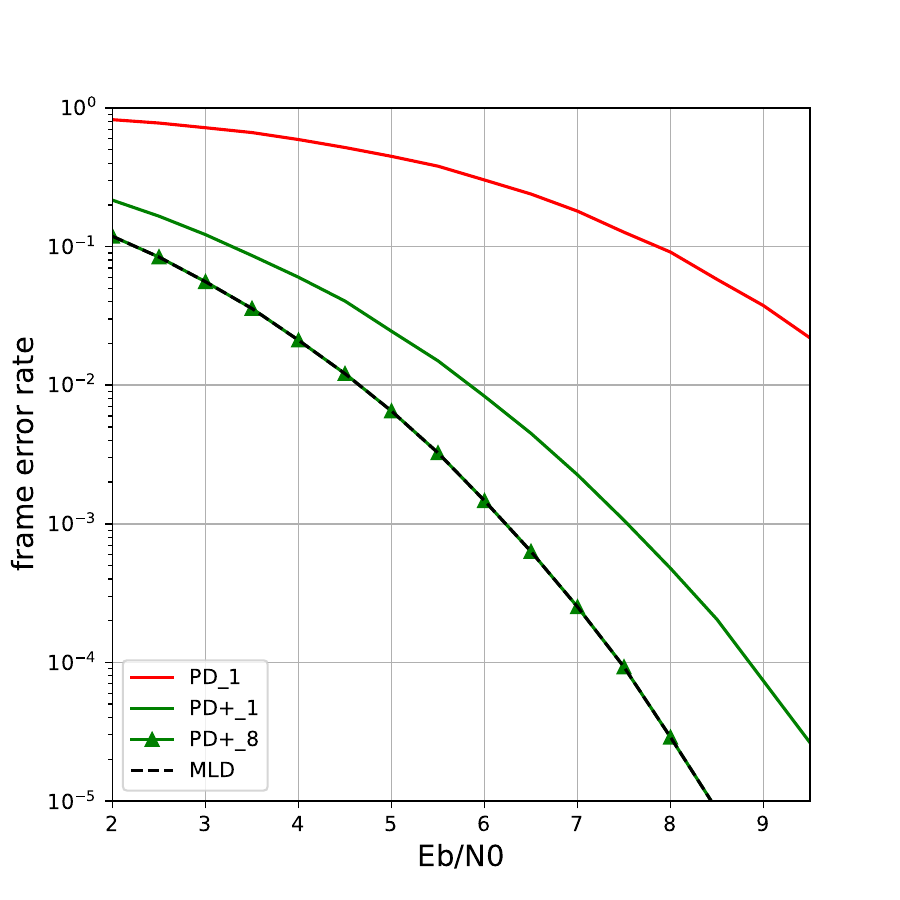}
\caption{Frame error rate vs $E_b/N_0$ for the $(16, 8)$ code generated by \eqref{eqn:RLC}.}
\label{fig:RLC}
\end{figure}
}

\section{Concluding Remark}\label{sec:conclude}
In this paper, building on the idea of $\pd$ presented in \cite{PD}, we propose a novel universal decoding scheme, $\pde$, for BLBCs. The core concept is to transform the BLBC being decoded into a polar-like code that can be efficiently decoded using existing polar code decoding algorithms. This target polar-like code is constructed by pruning the kernel of {\black Ar\i kan's} polar code and applying code shortening.

To achieve this transformation, we developed an AI-inspired search algorithm to identify a suitable target polar-like code. Extensive simulations were conducted to evaluate the effectiveness of the proposed approach. The results demonstrate that $\pde$ outperforms or is comparable to existing decoders, such as OSD and GRAND, while achieving significantly lower complexity. Notably, the forward compatibility of $\pde$ allows it to benefit from current and future advancements in polar code decoding algorithms.

Future work includes the following directions: 1) Extending $\pde$ to longer blocklengths: This presents a significant challenge due to the exponential growth of the search space with blocklength; 2) Generalizing $\pde$ to non-binary codes, such as {\black Reed--Solomon} codes; 3) Applying more powerful tools from AI and machine learning to solve the local search problem at hand.


\section*{Acknowledgment}
Yu-Chih Huang would like to thank Prof. Krishna R. Narayanan of Texas A\&M University for his insightful comments and fruitful discussions, as well as Prof. Shu Lin of the University of California, Davis, for his encouragement.

\bibliographystyle{IEEEtran}
\bibliography{journal_abbr,bib_polar_pn}

\begin{thebibliography}{10}
\providecommand{\url}[1]{#1}
\csname url@samestyle\endcsname
\providecommand{\newblock}{\relax}
\providecommand{\bibinfo}[2]{#2}
\providecommand{\BIBentrySTDinterwordspacing}{\spaceskip=0pt\relax}
\providecommand{\BIBentryALTinterwordstretchfactor}{4}
\providecommand{\BIBentryALTinterwordspacing}{\spaceskip=\fontdimen2\font plus
\BIBentryALTinterwordstretchfactor\fontdimen3\font minus
  \fontdimen4\font\relax}
\providecommand{\BIBforeignlanguage}[2]{{%
\expandafter\ifx\csname l@#1\endcsname\relax
\typeout{** WARNING: IEEEtran.bst: No hyphenation pattern has been}%
\typeout{** loaded for the language `#1'. Using the pattern for}%
\typeout{** the default language instead.}%
\else
\language=\csname l@#1\endcsname
\fi
#2}}
\providecommand{\BIBdecl}{\relax}
\BIBdecl

\bibitem{shulin09}
W.~E. Ryan and S.~Lin, \emph{Channel Codes - Classical and Modern}.\hskip 1em
  plus 0.5em minus 0.4em\relax Cambridge University Press, 2009.

\bibitem{Hamming1950}
R.~W. Hamming, ``{Error detecting and error correcting codes},'' \emph{Bell
  Syst. Tech. J}, vol.~29, no.~2, pp. 147--160, 1950.

\bibitem{Gallager62}
R.~G. Gallager, ``Low density parity-check codes,'' \emph{IRE Transactions on
  Information Theory}, vol.~8, no.~1, pp. 21--28, Jan. 1962.

\bibitem{arikan09}
E.~Arikan, ``Channel polarization: A method for constructing capacity-achieving
  codes for symmetric binary-input memoryless channels,'' \emph{IEEE
  Transactions on Information Theory}, vol.~55, no.~7, pp. 3051--3073, Jul.
  2009.

\bibitem{Mondelli16polar_floor}
M.~Mondelli, S.~H. Hassani, and R.~L. Urbanke, ``Unified scaling of polar
  codes: Error exponent, scaling exponent, moderate deviations, and error
  floors,'' \emph{IEEE Transactions on Information Theory}, vol.~62, no.~12,
  pp. 6698--6712, 2016.

\bibitem{5G_book18}
\BIBentryALTinterwordspacing
E.~Dahlman, S.~Parkvall, and J.~Skold, \emph{5G NR: The Next Generation
  Wireless Access Technology}.\hskip 1em plus 0.5em minus 0.4em\relax Academic
  Press, 2018. [Online]. Available:
  \url{https://books.google.com.tw/books?id=C5poDwAAQBAJ}
\BIBentrySTDinterwordspacing

\bibitem{Jinhong_6Gcodes}
M.~Rowshan, M.~Qiu, Y.~Xie, X.~Gu, and J.~Yuan, ``Channel coding toward {6G}:
  Technical overview and outlook,'' \emph{IEEE Open Journal of the
  Communications Society}, vol.~5, pp. 2585--2685, 2024.

\bibitem{blahut2003algebraic}
\BIBentryALTinterwordspacing
R.~Blahut, \emph{Algebraic Codes for Data Transmission}.\hskip 1em plus 0.5em
  minus 0.4em\relax Cambridge University Press, 2003. [Online]. Available:
  \url{https://books.google.com.tw/books?id=4fWUAwAAQBAJ}
\BIBentrySTDinterwordspacing

\bibitem{Krishna08}
J.~Jiang and K.~R. Narayanan, ``Algebraic soft-decision decoding of
  {Reed–Solomon} codes using bit-level soft information,'' \emph{IEEE
  Transactions on Information Theory}, vol.~54, no.~9, pp. 3907--3928, 2008.

\bibitem{Li_Chen22}
Y.~Wan, L.~Chen, and F.~Zhang, ``Algebraic soft decoding of elliptic codes,''
  \emph{IEEE Transactions on Communications}, vol.~70, no.~3, pp. 1522--1534,
  2022.

\bibitem{OSD}
M.~P.~C. Fossorier and S.~Lin, ``Soft-decision decoding of linear block codes
  based on ordered statistics,'' \emph{IEEE Transactions on Information
  Theory}, vol.~41, no.~5, pp. 1379--1396, Sep. 1995.

\bibitem{GRAND}
K.~R. Duffy, J.~Li, and M.~Médard, ``Capacity-achieving guessing random
  additive noise decoding,'' \emph{IEEE Transactions on Information Theory},
  vol.~65, no.~7, pp. 4023--4040, 2019.

\bibitem{ORBGRAND}
K.~R. Duffy, W.~An, and M.~Médard, ``Ordered reliability bits guessing random
  additive noise decoding,'' \emph{IEEE Transactions on Signal Processing},
  vol.~70, pp. 4528--4542, 2022.

\bibitem{PD}
C.-Y. Lin, Y.-C. Huang, S.-L. Shieh, and P.-N. Chen, ``Transformation of binary
  linear block codes to polar codes with dynamic frozen,'' \emph{IEEE Open
  Journal of the Communications Society}, vol.~1, pp. 333--341, 2020.

\bibitem{Trifonov13}
P.~Trifonov and V.~Miloslavskaya, ``Polar codes with dynamic frozen symbols and
  their decoding by directed search,'' in \emph{Proc. IEEE ITW}, Sep. 2013.

\bibitem{Fast_SCL}
G.~Sarkis, P.~Giard, A.~Vardy, C.~Thibeault, and W.~J. Gross, ``Fast list
  decoders for polar codes,'' \emph{IEEE Journal on Selected Areas in
  Communications}, vol.~34, no.~2, pp. 318--328, 2016.

\bibitem{Fast_flex_SCL}
S.~A. Hashemi, C.~Condo, and W.~J. Gross, ``Fast and flexible
  successive-cancellation list decoders for polar codes,'' \emph{IEEE
  Transactions on Signal Processing}, vol.~65, no.~21, pp. 5756--5769, 2017.

\bibitem{Memory_eff_SC}
S.~A. Hashemi, C.~Condo, F.~Ercan, and W.~J. Gross, ``Memory-efficient polar
  decoders,'' \emph{IEEE Journal on Emerging and Selected Topics in Circuits
  and Systems}, vol.~7, no.~4, pp. 604--615, 2017.

\bibitem{Rowshan22rewind}
M.~Rowshan and E.~Viterbo, ``Efficient partial rewind of successive
  cancellation-based decoders for polar codes,'' \emph{IEEE Transactions on
  Communications}, vol.~70, no.~11, pp. 7160--7168, 2022.

\bibitem{Golay2412}
X.-H. Peng and P.~Farrell, ``On construction of the (24, 12, 8) {Golay}
  codes,'' \emph{IEEE Transactions on Information Theory}, vol.~52, no.~8, pp.
  3669 -- 3675, Aug. 2006.

\bibitem{Tal15}
I.~Tal and A.~Vardy, ``List decoding of polar codes,'' \emph{IEEE Transactions
  on Information Theory}, vol.~61, no.~5, pp. 2213--2226, May 2015.

\bibitem{MultiKernel}
F.~Gabry, V.~Bioglio, I.~Land, and J.-C. Belfiore, ``Multi-kernel construction
  of polar codes,'' in \emph{IEEE ICC}, May 2017.

\bibitem{simu_anneal}
P.~van Laarhoven and E.~Aarts, \emph{Simulated Annealing: Theory and
  Applications}.\hskip 1em plus 0.5em minus 0.4em\relax Berlin, Germany:
  Springer Science and Business Media, 1987.

\bibitem{viterbo_SCLflip}
M.~Rowshan and E.~Viterbo, ``{SC} list-flip decoding of polar codes by shifted
  pruning: A general approach,'' \emph{Entropy}, vol.~24, no.~9, 2022.

\bibitem{BP_polar}
E.~Arikan, ``Polar codes: A pipelined implementation,'' in \emph{Proc. 4th
  ISBC}, 2010.

\bibitem{BP_list}
A.~Elkelesh, M.~Ebada, S.~Cammerer, and S.~ten Brink, ``Belief propagation list
  decoding of polar codes,'' \emph{IEEE Communications Letters}, vol.~22,
  no.~8, pp. 1536--1539, 2018.

\bibitem{Seq_dec_polar}
V.~Miloslavskaya and P.~Trifonov, ``Sequential decoding of polar codes,''
  \emph{IEEE Communications Letters}, vol.~18, no.~7, pp. 1127--1130, 2014.

\bibitem{Automorphism_Ensemble_Dec}
V.~Bioglio, I.~Land, and C.~Pillet, ``Group properties of polar codes for
  automorphism ensemble decoding,'' \emph{IEEE Transactions on Information
  Theory}, vol.~69, no.~6, pp. 3731--3747, 2023.

\bibitem{PeihongYuan24}
P.~Yuan and M.~C. Coşkun, ``Successive cancellation ordered search decoding of
  modified {$G_N$}-coset codes,'' \emph{IEEE Transactions on Communications},
  vol.~72, no.~6, pp. 3141--3154, 2024.

\bibitem{ShuLin_kernel}
H.-P. Lin, S.~Lin, and K.~A.~S. Abdel-Ghaffar, ``Linear and nonlinear binary
  kernels of polar codes of small dimensions with maximum exponents,''
  \emph{IEEE Transactions on Information Theory}, vol.~61, no.~10, pp.
  5253--5270, 2015.

\bibitem{viterbo20kernel}
F.~Abbasi and E.~Viterbo, ``Large kernel polar codes with efficient window
  decoding,'' \emph{IEEE Transactions on Vehicular Technology}, vol.~69,
  no.~11, pp. 14\,031--14\,036, 2020.

\bibitem{Wang_prune21}
H.-P. Wang and I.~M. Duursma, ``Log-logarithmic time pruned polar coding,''
  \emph{IEEE Transactions on Information Theory}, vol.~67, no.~3, pp.
  1509--1521, 2021.

\bibitem{PeihongYuanConjecture}
P.~Yuan, T.~Prinz, G.~Boecherer, O.~Iscan, R.~Boehnke, and W.~Xu, ``Polar code
  construction for list decoding,'' in \emph{SCC 2019; 12th International ITG
  Conference on Systems, Communications and Coding}, 2019, pp. 1--6.

\bibitem{Rowshan23GRAND}
M.~Rowshan and J.~Yuan, ``Low-complexity {GRAND} by segmentation,'' in
  \emph{Proc. IEEE GLOBECOM}, 2023, pp. 6145--6151.

\bibitem{QR18}
Y.~Li, Y.~Duan, H.-C. Chang, H.~Liu, and T.-K. Truong, ``Using the difference
  of syndromes to decode quadratic residue codes,'' \emph{IEEE Transactions on
  Information Theory}, vol.~64, no.~7, pp. 5179--5190, 2018.

\bibitem{QR20}
Y.~Duan and Y.~Li, ``An improved decoding algorithm to decode quadratic residue
  codes based on the difference of syndromes,'' \emph{IEEE Transactions on
  Information Theory}, vol.~66, no.~10, pp. 5995--6000, 2020.

\end{thebibliography}

\end{document}